\documentclass[twocolumn,showpacs,amsmath,amssymb,nofootinbib]{revtex4-1}
\usepackage{dcolumn}
\usepackage{bm}
\usepackage{graphicx}

\newcommand{\be}{\begin{equation}}
\newcommand{\ee}{\end{equation}}
\newcommand{\ba}{\begin{eqnarray}}
\newcommand{\ea}{\end{eqnarray}}
\newcommand{\nn}{\nonumber\\}
\newcommand{\n}[1]{\label{#1}}
\newcommand{\eq}[1]{(\ref{#1})}

\newcommand{\hh}{\, ,\hspace{0.5cm}}
\newcommand{\hhh}{\, ,\hspace{0.2cm}}

\newcommand{\BM}[1]{{\mbox{\boldmath $#1$}}}

\newcommand{\ve}{\varepsilon}

\newcommand{\bi}[1]{\bibitem{#1}}

\newcommand{\PRD}[4]{{#1}{Phys. Rev.\  D\ }{\bf #2},\  #3 (#4).}
\newcommand{\PRA}[4]{{#1}{Phys. Rev.\  A\ }{\bf #2},\  #3 (#4).}

\newcommand{\PR}[4]{{#1}{Phys. Rev.\ }{\bf #2},\  #3 (#4).}

\begin{document}

\title{Scattering of circularly polarized light by a rotating black hole}
\author{Valeri P. Frolov}
\email{vfrolov@ualberta.ca}
\author{Andrey A. Shoom}
\email{ashoom@ualberta.ca}
\affiliation{Theoretical Physics Institute, University of Alberta,
Edmonton, AB, Canada,  T6G 2E1}

\begin{abstract}
We study scattering of polarized light by a rotating (Kerr) black hole of the mass $M$ and the angular momentum $J$. In order to keep trace of the polarization dependence of photon trajectories one can use the following dimensionless parameter: $\ve=\pm (\omega M)^{-1}$, where $\omega$ is the photon frequency and the sign $+$ ($-$) corresponds to the right (left) circular polarization. 
 We assume that $|\ve| \ll 1$ and use the modified geometric optics approximation developed in \cite{FS}, that is we include the first order in $\ve$ polarization dependent terms into the eikonal equation. These corrections modify late time behavior of photons. We demonstrate that the photon moves along a null curve, which in the limit $\ve=0$ becomes a null geodesic. We focus on the scattering problem for polarized light.
 Namely, we consider the following problems: (i) How does the photon bending angle depend on its polarization; (ii) How does position of the image of a point-like source depend on its polarization; (iii) How does the arrival time of photons depend on their polarization. We perform the numerical calculations that illustrate these effects for an extremely rotating black hole and discuss their possible applications.
\end{abstract}

\pacs{04.70.Bw, 04.25.-g, 42.15.Dp \hfill Alberta-Thy-9-12}

\maketitle

\section{Introduction}

There exists a well-known analogy between gravity and electromagnetism. In particular, the linearized Einstein equations can be written in the form similar to the Maxwell equations (see, e.g., the  review \cite{Mashhoon_08} and references therein). Analyzing  propagation of a circularly polarized light in a rotating frame in a flat spacetime, Mashhoon \cite{Mashhoon_93} demonstrated that there exists helicity-rotation coupling with the energy proportional to $\pm \hbar (\BM{k},\BM{\Omega})$, where $\BM{k}$ is the wave vector of a photon, and $\BM{\Omega}$ is the angular velocity of the rotation. This effect is similar to energy splitting of a particle with magnetic moment in a magnetic field. When such a particle moves in an inhomogeneous magnetic field its trajectory depends on the spin orientation. One can expect that there exists a gravitational analogue of this Stern-Gerlach effect. Namely, in the gravitational field of a rotating body trajectories of circularly polarized photons depend on their polarization. Using this analogy, Mashhoon \cite{,Mashhoon_93,Ma:74a,Ma:75} demonstrated
that photons of the opposite circular polarization emitted by a distant source, after scattering, deflect to directions with the separation angle,
\be\n{sepang}
\mbox{separation angle}\sim {\lambda GJ\over c^3 D^3}\, .
\ee
Here $\lambda$ is the wavelength of the photon, $J$ is the angular momentum of the rotating body, and $D$ is the characteristic distance from the photon to the body at the moment of their minimal separation. Mashhoon also arrived to the conclusion that in order to describe these effects in the geometric optics approximation, the corresponding corrections should be included in the eikonal approximation. It should be emphasized that these conclusions are based on the gravitational Larmor theorem \cite{Mashhoon_93} and gravito--electromagnetic analogy, valid in the weak field approximation.

One can expect that the dependence of the photon trajectory on its
helicity  might be enhanced in the case when the photon moves in a strong gravitational field, for example close to a rapidly rotating black hole. In order to derive this dependence, one needs to modify the standard geometric optics (GO) approximation and adapt it to the propagation of circularly polarized beams of light. The standard GO scheme is well known (see, e.g., \cite{MTW}). In this approach one starts with the following anzats for the vector potential of the electromagnetic field in the Lorentz gauge:
\be\n{anz}
A_{\mu}\sim \Re(a_{\mu} e^{iS/\epsilon})\, .
\ee
One assumes that $\epsilon$ is a small parameter, so that the wavelength of light $\lambda$ is much smaller than the other scales $L$, $\epsilon=\lambda/L\ll 1$.
After a substitution of this anzats into the Maxwell equations
\be
\Box A_{\mu}-R_{\mu}^{\ \nu} A_{\nu}=0\, ,
\ee
one obtains in the lowest order approximation  the following eikonal equation:
\be\n{ei}
(\nabla S)^2=0\, .
\ee
This first order partial differential equation can be solved by the method of characteristics by identifying $\nabla_{\mu}S=k_{\mu}$, where $k^{\mu}$ is a vector tangent to null geodesics describing the photon trajectory. The next order in $\epsilon$ equations show that the vector of linear polarization is parallel transported along the worldline of the photon and the following conservation law is satisfied:
\be
\nabla_{\mu}(\BM{a}^2 k^{\mu})=0\, .
\ee
Higher order corrections give expansion of the amplitude $a_{\mu}$ in terms of the small parameter $\epsilon$, while the eikonal function $S$ is kept unchanged.

For a real scalar field the anzats similar to \eq{anz} is well defined and unique if one chose the prefactor to be real. For a many component field there is an ambiguity: one can add to $S$ higher order `correction' $\epsilon\varphi$ and, at the same time,  change the prefactor $a_{\mu}\to \exp(-i\varphi)a_{\mu}$ (for discussion of the WKB and GO approximations for multi-component field see, e.g., \cite{weinberg}). A similar problem arises, for example, when one considers WKB approximation for the Dirac equation describing an electron moving in the external electromagnetic field (see, e.g., \cite{BjoOrb,WeiLit,KraNaiFuj,KraBie,BolKep,Kep,Bol,Duv}). The recent nice paper \cite{HBH}  clearly illustrates problems that arise when one tries to adapt WKB approximation for the description of the Stern-Gerlach effect. It is shown that the correct answer is obtained when one first diagonalizes the corresponding Pauli equation, and then includes the spin corrections, which are of the first order in $\hbar$, into the eikonal equation. Namely, these first order terms modify the propagation of the electron at late time (at far distance) and allow one to obtain the correct result. This result directly follows from the stationary phase approximation for the path integral for the electron in the magnetic field. As a result of this approximation, one restores the theory of the electron as a particle with `a classical spin', and the corresponding Dirac (Pauli) equation is nothing but the first quantized theory.

Unfortunately, a similar approach does not work for an electromagnetic field. The problems are basically connected with gauge invariance. A massless photon can be easily defined in the momentum space, while it is a non-trivial problem to define a localized classical photon in the configuration space (in the spacetime). The comprehensive discussion of this problem can be found in \cite{OK}, see also the references therein.

In our paper \cite{FS}, in order to study propagation of circularly polarized monochromatic electromagnetic waves in a stationary spacetime, we used another approach. Using the projector along the direction of the Killing vector, the Maxwell equations are reduced to $(3+1)-$dimensional form. The three-dimensional complex vectors $\BM{F}^{\pm}=\BM{E}\pm i\BM{H}$, that are linear combinations of the electric $\BM{E}$ and magnetic $\BM{H}$ fields, are used to describe fields with fixed circular polarization. In a flat spacetime $\BM{F}^{\pm}$ are known as the Riemann-Silberstein vectors (see, e.g., \cite{BB}). It is easy to show that in a stationary spacetime their  positive frequency Fourier transform $\BM{ {\cal F}}^{\pm}$ obeys a single master equation, which describes a state with a given circular polarization.

In \cite{FS}, we used the GO anzats for $\BM{ {\cal F}}^{\pm}$,
\be
\BM{ {\cal F}}^{\pm}=\BM{f}^{\pm}e^{i\tilde{S}^{\pm}/\epsilon}\, .
\ee
For monochromatic waves of high frequency $\omega$,  $\epsilon\sim \omega^{-1}$ is a small parameter.
We demonstrated that there exists a natural prescription for the choice of the eikonal function. According to this prescription, the lowest order helicity dependent corrections should be included in  $\tilde{S}^{\pm}$, while the transport equation for the polarization vector $\BM{f}^{\pm}$ (at least in the lowest order in $\epsilon$) does not depend on the state of polarization. In such a modified GO approach the eikonal equation, which determines the effective Hamiltonian for light rays, includes terms proportional to $\epsilon$. The sign of this terms depends on the helicity. As a result, the modified Hamiltonian describes rays with worldlines depending on the polarization of the photon. For short time asymptotics such a modification may be considered as a simple re-arrangement of terms in the asymptotic expansion in the parameter $\epsilon$, while in the late time regime it results in the spatial separation of the light rays for photons with opposite circular polarization. For initially linearly polarized light at late time one obtains a mixture of two opposite circularly polarized photons, which are moving along spatially different trajectories. 

In a stationary spacetime this effect exists only when the gravitating body is rotating. If $t$ is the Killing time, then for a rotating mass the gravitational field contains components $g_{ti}$ ($i=1,2,3$) which cannot be banished by the spatial coordinate transformation $x^i\to \tilde{x}^i(x^i)$. The remarkable fact is that additional corrections to the 3D equations of motion of the photon connected with its polarization can be described as the additional force $\pm \epsilon\,\mbox{curl}\,\mbox{curl}\,\BM{g}$ in these equations \cite{FS}.

The goal of this paper is to study the scattering of a circularly polarized light by a rotating (Kerr) black hole.
The paper is organized as follows. In Sec. II we introduce ultrastationary metric for stationary spacetime and discuss $(3+1)-$form of the polarized ray equations. We demonstrate that circularly polarized beams of light propagate along null rays, which become null geodesics in the limit $\epsilon=0$ (that is, for unpolarized light). We also describe some general results concerning time of arrival of polarized photons emitted at some point to a distant observer. In Sec. III we apply this formalism to the Kerr spacetime. In addition, we introduce the advanced and retarded star time coordinates which increase monotonically along the null rays and which are convenient for their parametrization. We present the equations for the propagation of circularly polarized photons in dimensionless form which is suitable for numerical integration. Section IV contains formulation of the scattering problem in the Kerr spacetime adapted to the circularly polarized photons. In Sec. V we study how polarization affects bending angles of scattered photons, and in Sec. VI we study how position of the image of a photon and its time of arrival depend on polarization. We discuss the results in Sec. VII. Details of the numerical computations are given in the Appendix.  

In the present paper we use the system of units where $G=c=1$ and the sign conventions for the metric
and other geometrical quantities adopted in the book  \cite{MTW}.

\section{Circularly polarized light propagation in a stationary spacetime}

\subsection{Effective equation for polarized photons}

Let us consider the following equation in a curved spacetime with the metric $g_{\mu\nu}$:
\be\n{eq1}
{D^2x^{\mu}\over d\lambda^2}=P {dx^{\mu}\over d\lambda} +\ve g^{\mu\rho}F_{\rho\nu}{dx^{\nu}\over d\lambda}\, ,
\ee
where $D/d\lambda$ is the covariant derivative, $P$ is a scalar function, $\ve$ is a small parameter, and  $F_{\rho\nu}=F_{[\rho\nu]}$. We shall demonstrate later that the equation for propagation of circularly polarized light in a stationary spacetime is of this form. Let us at first discuss general properties of this equation. Let us multiply Eq. \eq{eq1} by $g_{\mu\kappa}dx^{\kappa}/d\lambda$, then one has
\be
{dV\over d \lambda}=2PV\hh
V=g_{\mu\kappa}{dx^{\mu}\over d\lambda} {dx^{\kappa}\over d\lambda}\,.
\ee
This relation implies that if for some value of $\lambda$ the function $V$ vanishes, then $V=0$ along the complete worldline, and hence this line is null. The reparametrization
\be\n{rep}
d\lambda=\beta d\tilde{\lambda}\,,
\ee
where $\beta$ is a scalar function, preserves the form of Eq. \eq{eq1}, and one has
\ba
{D^2x^{\mu}\over d\tilde{\lambda}^2}&=&\tilde{P}{dx^{\mu}\over d\tilde{\lambda}} +\ve\beta g^{\mu\rho}F_{\rho\nu}{dx^{\nu}\over d\tilde{\lambda}}\,,\n{eqp}\\
\tilde{P}&=&\left(\beta P +{d\ln\beta\over d\tilde{\lambda}}\right)\,.
\ea
Thus, by proper rescaling one can always put $\tilde{P}=0$. We call such a parametrization, defined up to a linear rescaling, an affine parametrization. For $\ve=0$, Eq. \eq{eq1} is a geodesic equation. For null geodesics the canonical parameter $\lambda$ is a usual affine parameter.

Consider now a conformal transformation of the metric,
\be
g_{\mu\nu}=\Omega^2 \tilde{g}_{\mu\nu}\, .
\ee
For null worldlines Eq. \eq{eq1} is transformed as follows:
\be\n{eqc}
{\tilde{D}^2x^{\mu}\over d\lambda^2}=\left(P-2{d\ln{\Omega}\over d\lambda}\right)  {dx^{\mu}\over d\lambda} +\ve \Omega^{-2}\tilde{g}^{\mu\rho}F_{\rho\nu}{dx^{\nu}\over d\lambda}\, .
\ee
By comparing this result with Eq. \eq{eqp} it is easy to see that the canonical form of Eq. \eq{eq1},
\be\n{eqcan}
{D^2x^{\mu}\over d\lambda^2}=\ve g^{\mu\rho}F_{\rho\nu}{dx^{\nu}\over d\lambda}\, ,
\ee
is conformal invariant, provided under this transformation $\tilde{F}_{\mu\nu}={F}_{\mu\nu}$, and the conformal transformation is accompanied by the reparametrization \eq{rep} with $\beta=\Omega^2.$\footnote{It is interesting to notice that Eq. \eq{eqcan} can be obtained as a limiting case of the equation of motion of a charged particle of the mass $m$ and the electric charge $e$. This equation is of the form
\be\n{a1}
m{D^2x^{\mu}\over d\tau^2}=e F^{\mu}_{\ \ \nu}{dx^{\nu}\over d\tau}\, ,\nonumber
\ee
where $D/d\tau$ is the covariant derivative corresponding to the metric $g_{\mu\nu}$.
Denote by $\lambda$ a parameter connected with the proper time $\tau$ as follows:
\be
\tau=m\lambda\,.\nonumber
\ee
Then, this equation takes the form
\be
\ddot{x}^{\mu}=eF^{\mu}_{\ \ \nu}\dot{x}^{\nu}\, ,\nonumber
\ee
where the over dot stands for the covariant derivative $D/d\lambda$. It is easy to check that
\be\n{a3}
\dot{x}^2\equiv \dot{x}^{\mu}\dot{x}_{\mu}=-m^2\, .\nonumber
\ee
The above equations allow for the ultrarelativistic limit $m\to 0$, and in this limit one obtains Eq. \eq{eqcan} with $e=\ve$.}

\subsection{Stationary spacetime and ultrastationary metric}

To study how polarization affects the light propagation in a stationary spacetime, it is convenient to use what we call ultrastationary form of the metric. Let us discuss the related formalism. We present the metric $ds^2$ of a stationary spacetime  in the form
\ba
ds^2&=&{g}_{\mu\nu}dx^{\mu}dx^{\nu}=h\,dS^2\hhh h=-g_{tt}\equiv -\xi^2_{(t)}\,,\\
dS^2&=&-(dt-g_i\,dx^i)^2+dl^2\hhh g_i=g_{t i}/h ,\\
dl^2&=&\gamma_{ij}\,dx^i\,dx^j\hhh \gamma_{ij}=h^{-1} g_{ij}+g_i g_j\, .
\ea
Here $\xi^\mu_{(t)}=\delta^{\mu}_{t}$ is a timelike at infinity Killing vector. For $g_i=0$ the metric $dS^2$ is ultrastatic. We call the metric $dS^2$ {\em ultrastationary}.

Source-free Maxwell equations in 4D are conformally invariant. For this reason, in order to study their solutions, one can use either the original spacetime metric $ds^2$ or its ultrastationary version $dS^2$. Similarly, a null ray in the physical metric $ds^2$ is at the same time a null ray in the conformal metric  $dS^2$. Moreover, after the conformal transformation, a null geodesic remains a null geodesic. However, its affine parametrization depends on the conformal factor. In our discussion we shall use the ultrastationary form of the metric, that simplifies essentially many relations.

\subsection{$(3+1)$ form of the equations}

Because of the conformal invariance of a null ray solutions to the effective equation \eq{eqcan}, it is sufficient to study their properties in the ultrastationary spacetime. Consider a curve $x^{\mu}(\lambda)$ and denote by $u^{\mu}=\dot{x}^{\mu}\equiv dx^{\mu}/d\lambda$ its tangent vector. We also use the notation
\be
(\ldots)^{.}=u^{\mu}{\nabla}_{\mu}(\ldots)\, ,
\ee
where ${\nabla}_{\mu}$ is the covariant derivative in the ultrastationary metric $dS^{2}$. In these notations the effective equation \eq{eqcan} takes the form
\be\n{a2}
\ddot{x}^{\mu}=\ve F^{\mu}_{\ \ \nu}\dot{x}^{\nu}\, .
\ee
Let us write this equation in a $(3+1)$ form constructed by projecting it along the Killing vector $\xi^{\mu}_{(t)}$.
We notice that the only non-vanishing components of the Christoffel symbols with the only one index $t$ are
\be
\Gamma_{tij}={1\over 2}(g_{i,j}+g_{j,i})\hhh\Gamma_{ijt}=\frac{1}{2}(g_{i,j}-g_{j,i})\, .
\ee
Denote by ${\cal G}_{ijk}$ the Christoffell symbols for the 3D metric $\gamma_{ij}$. Then one has
\ba\n{a8}
\Gamma_{ijk}&=&{\cal G}_{ijk}+{1\over 2}\left[ g_{j}(g_{k,i}-g_{i,k})\right.\nonumber\\
&+& \left. g_{k}(g_{j,i}-g_{i,j})-g_{i}(g_{j,k}+g_{k,j})\right]\,.
\ea
Denote $\ddot{x}_{\mu}\equiv \tilde{g}_{\mu\nu}\ddot{x}^{\nu}$, where $\tilde{g}_{\mu\nu}$ is the ultrastationary metric.  Then one has
\be\n{a5}
\ddot{x}_{t}\equiv -{d^2 t\over d\lambda^2}+g_i {d^2 x^i\over d\lambda^2}+g_{i,j}{d x^i\over d\lambda}{d x^j\over d\lambda}\, .
\ee
Denote
\be\n{a6}
U\equiv{d t\over d\lambda}-g_i {d x^i\over d\lambda}\, ,
\ee
then Eq. \eq{a5} can be written in the form
\be\n{a7}
\ddot{x}_{t}=-{dU\over d\lambda}\, .
\ee
Using this relation and Eq. \eq{a6} one obtains
\be\n{a9}
\ddot{x}_{i}=\gamma_{ij}{D^2 x^j\over d\lambda^2}+U(g_{i,j}-g_{j,i}){dx^j\over d\lambda}+g_{i}\frac{dU}{d\lambda}\, ,
\ee
where $D/d\lambda$ is the covariant derivative corresponding to the 3D metric $\gamma_{ij}$.

Let us consider a special case of the field $F_{\mu\nu}$ when $F_{ti}=0$. Then the 4D equation \eq{eqcan} has the following 3D form:
\ba
&&\frac{dU}{d\lambda}=0\hh \gamma_{ij}{D^2 x^j\over d\lambda^2}={\cal F}_{ij}{dx^j\over d\lambda}\, ,\n{a10}\\
&&{dt\over d\lambda}=U+g_{k}{dx^k\over d\lambda}\, ,\n{a11}\\
&&{\cal F}_{ij}=-U(g_{i,j}-g_{j,i})+\ve F_{ij}\, .\n{a12}
\ea
Here $U$ is the constant of motion. By the rescaling $\lambda\to A\lambda$ ($A=const$) one can always put $U=1$. In what follows, we shall use this choice.

Denote $\BM{x}=(x^1,x^2,x^3)$, then Eqs. \eq{a10}-\eq{a12} can be written in the following 3D form:
\ba
{D^2\BM{x}\over dl^2}&=&\left[\frac{d\BM{x}}{dl}\times(\mbox{curl}\,\BM{g}+\ve \BM{F})\right]\,,\n{B2}\\
\frac{dt}{dl}&=&1+\left(\BM{g},\frac{d\BM{x}}{dl}\right)\,.\n{B3}
\ea
Here the covariant derivative $D/dl$ is defined with respect to the metric $\gamma_{ij}$. In what follows, all the operations involving the Latin indices $i,j, \ldots$ are performed by using the metric $\gamma_{ij}$ and its inverse $\gamma^{ij}$. In particular, the vector and the scalar products of vectors $\BM{a}$ and $\BM{b}$ are given by
\be
(\BM{a},\BM{b})=a^{i}\gamma_{ij}b^{j}\hhh [\BM{a}\times\BM{b}]^{i}=e^{ijk}a_{j}b_{k}\hhh e^{ijk}=\frac{\epsilon^{ijk}}{\sqrt{\gamma}}\,,
\ee
respectively. Here $\epsilon^{ijk}$ is the three-dimensional completely antisymmetric Levi-Civita symbol normalized by the condition $\epsilon^{123}=1$.

It was shown in \cite{FS} that the equations for propagation of circularly polarized monochromatic light of the frequency $\omega$ have the form \eq{B2}-\eq{B3} with
\ba\n{fff}
F_{ij}&=&2\Phi_{[i,j]}\hhh\Phi^{i}=-(\mbox{curl}\,\BM{g})^{i}=[\nabla\times\BM{g}]^{i}=e^{ijk}g_{j,k}\,,\nn
\BM{F}&=&\mbox{curl}\,\mbox{curl}\,\BM{g}\, .
\ea
One also has $\ve=\pm(2\omega M)^{-1}$, and the signs $(-)$ and $(+)$ define the left and the right circular polarization, respectively. In what follows, we use the subscript $(0)$ for a null geodesic, while the subscripts $(-)$ and $(+)$ stand for the left and right polarization of photons, respectively.

\subsection{Time delay and Fermat principle}

Suppose $\Gamma$ is a timelike worldline of an observer and $t$ is the proper time along it. Suppose at some moment of time $t_o$ the observer registered light emitted at some point $e$. Consider also null rays emitted from the same point $e$ and obeying Eq. \eq{eqcan}. We assume that for each of the value of the parameter $\ve$ from a small interval there exists a null curve intersecting $\Gamma$. Let us denote by $t(\ve)$ the time of its arrival to $\Gamma$. Thus, one has a one-parametric family of null rays emitted at $e$ and intersecting $\Gamma$. It was proved in the papers \cite{NS,Perl1,Perl2} that $(dt(\ve)/d\ve)|_{\ve=0}=0$, and if there is no conjugated points along the null geodesic, then $t(0)$ is minimum of $t(\ve)$. In other words, in the general case the null geodesic arrives to the observer $\Gamma$ earlier than other null curves.

This general result concerning the time delay of null rays takes a simple form in a stationary spacetime. Namely, let $\Gamma_{e}$ and $\Gamma_{o}$ be two timelike Killing trajectories, and $\BM{x}_{e}$ and $\BM{x}_{o}$ be two spatial points representing these trajectories in the 3D space. Let a photon emitted at $\BM{x}_{e}$ at the time $t_{e}$ is registered by an observer at $\BM{x}_{o}$ at the time $t_{o}$. The relation \eq{B3} shows that
\be\n{ttt}
t(\ve)=t_{e}+\int_{\BM{x}_e}^{\BM{x}_o} (dl +g_i dx^i)\, .
\ee
It is possible to show (see, e.g., \cite{Lan,Brill}) that for a one-parametric family of null curves close to a null geodesic, the latter gives the extremum of the arrival time $t(\ve)$. In the general case, this extremality  means that a photon moving along such a geodesic arrives to the point of observation earlier than any other photon which follow non-geodesic motion. Possible exception is the case when there exist more than one null geodesics connecting the emitter and the observer. This is nothing but the Fermat principle for light rays in a stationary gravitational field.

\section{Propagation of circularly polarized light in the Kerr spacetime}

\subsection{The Kerr spacetime}

The Kerr metric in the Boyer-Lindquist coordinates $x^{\alpha}=(t,r,\theta,\phi)$  can be presented in the form
\be\n{k2}
ds^2=h[-(dt-g_i\,dx^i)^2+dl^{2}] \,,
\ee
where
\ba
&&\hspace{-0.9cm}h=1-\frac{2Mr}{\Sigma}\hhh g_{i}=-\frac{2aMr}{\Sigma h}\sin^{2}\theta\delta_{i}^{~\phi}\,,\\
&&\hspace{-0.9cm}dl^{2}=\gamma_{ij}dx^{i}dx^{j}=\frac{\Sigma}{\Delta h}dr^{2}+\frac{\Sigma}{h}d\theta^{2}+\frac{\Delta\sin^{2}\theta}{h^{2}} d\phi^{2}\,,
\ea
and
\be
\Sigma=r^{2}+a^{2}\cos^{2}\theta\hhh \Delta=r^{2}-2Mr+a^{2}\,.
\ee
This metric describes the gravitational field of a black hole of the mass $M$ and the
angular momentum $J=aM$ ($a>0$, $a/M\leq1$).

The Kerr metric has two Killing vectors $\xi^{\mu}_{(t)}$ and $\xi^{\mu}_{(\phi)}$ and the second rank Killing tensor $K_{\mu\nu}$ ($K_{(\mu\nu;\alpha)}=0$). For a geodesic line $x^{\mu}(\lambda)$ with an affine parameter $\lambda$ ($u^{\mu}=dx^{\mu}/d\lambda$)
the corresponding conserved quantities are (see, e.g., \cite{MTW})
\ba
E&\equiv&-\xi^{\alpha}_{(t)}u_{\alpha}=-u_{t}=h\left({dt\over d\lambda}-g_{\phi}{d\phi\over d\lambda}\right)\,,\n{13}\\
L&\equiv&\xi^{\alpha}_{(\phi)}u_{\alpha}=u_{\phi}=h\left(g_{\phi}{dt\over d\lambda}+(\gamma_{\phi\phi}-g_{\phi}^{2}){d\phi\over d\lambda}\right)\,,\\
Q&\equiv&K-(L-aE)^{2}\hhh K=K^{\alpha\beta}u_{\alpha}u_{\beta}\,.\n{15}
\ea
Here $E$ is photon energy, $L$ is photon azimuthal angular momentum, and $Q$ is the Carter constant. These integrals of motion make the geodesic equation
\be
\frac{D^{2}x^{\mu}}{d\lambda^{2}}=0\,
\ee
completely integrable. In particular, this means that the geodesic equation can be written in the following first order form:
\ba
\Sigma\,\frac{dt}{d\lambda}&=&\frac{(r^{2}+a^{2})^{2}E-2aMrL}{\Delta}-a^{2}E\sin^{2}\theta\,,\n{11a}\\
\Sigma\,\frac{dr}{d\lambda}&=&\sigma_{r}\sqrt{R}\hhh \sigma_{r}=\pm1\,,\n{11b}\\
\Sigma\,\frac{d\theta}{d\lambda}&=&\sigma_{\theta}\sqrt{\Theta}\hhh \sigma_{\theta}=\pm1\,,\n{11c}\\
\Sigma\,\frac{d\phi}{d\lambda}&=&\frac{L}{\sin^{2}\theta}+\frac{2aMrE-a^{2}L}{\Delta}\,.\n{11d}
\ea
For photons the functions $R$ and $\Theta$ are
\ba
R&=&[(r^{2}+a^{2})E-aL]^{2}-\Delta[Q+(L-aE)^{2}]\,,\nn
\Theta&=&Q+a^{2}E^{2}\cos^{2}\theta-L^{2}\cot^{2}\theta\,.\n{11e}
\ea
Important property of a null geodesics of a Kerr spacetime is that it has no more than one radial turning point ( see, e.g., \cite{FZ}).

\subsection{Advanced and retarded star time}

In the study of the Kerr spacetime one often uses the retarded and advanced time coordinates
\be
u=t-r_*\hh v=t+r_*\hh r_*=\int{(r^2+a^2)dr\over \Delta}\, .
\ee
They are well-known generalization of the outgoing and ingoing  Eddington-Finkelstein coordinates of the Schwarzschild spacetime (see, e.g., \cite{MTW}). The latter are null coordinates, that is, surfaces $u=const$ and $v=const$ are null. However, as it was demonstrated in \cite{FW}, the coordinates $u$ and $v$ in the Kerr spacetime do not have this property. Namely, in the presence of rotation, surfaces $u=const$ and $v=const$ are timelike, and they become null only at infinity. For this reason, one cannot guarantee that $u$ and $v$ are monotonically increasing functions along any null curve.

However, it is possible to  introduce new coordinates which are almost everywhere timelike and which can be used as monotinic parameters along null curves. We define these coordinates as follows:
\ba\n{w}
W_{\pm}&\equiv& t\mp r_{\star}\hhh r_{\star}=\int\frac{H}{\Delta}dr\,,\\
H&=&(r^{4}+2a^{2}Mr+a^{2}r^{2})^{1/2}\,.
\ea
It is easy to see that
\be\n{50}
v-W_-=W_+-u=r_*-r_{\star}\,, \ \
dr_*-dr_{\star}={r^2+a^2-H\over \Delta}dr\, .
\ee
To distinguish $r_{\star}$ from the tortoise coordinate $r_*$, we call it {\em radial star coordinate}.

Expanding the last expression in Eq. \eq{50} at $r\to\infty$ one obtains
\be
dr_*-dr_{\star}\sim  {a^2\over 2r^2}\left( 1-{3\over 4}{a^2\over r^2}
- {1\over 2} {a^2 M\over r^3}+\ldots\right)dr\, .
\ee
This means that at a large distance the difference between $W_{-}$ and $W_{+}$ and the corresponding values of $v$ and $u$ is of the order of $a^2/r$, so that the new coordinates $W_{-}$ and $W_{+}$ are also asymptotically null.

Let us show that a surface $W_{\pm}=const$ is always non-timelike, so that $W_{\pm}$ can be used as a time coordinate. This property differs  $W_{\pm}$ from $u$ and $v$. Indeed, the gradient covector $W_{\pm;\mu}$ in the Boyer-Lindquist coordinates reads
\be
W_{\pm,\mu}=\left(1,\mp\frac{H}{\Delta},0,0\right)\, .
\ee
Using the expressions for the components of $g^{\mu\nu}$,
\be
g^{tt}={a^2\Delta\sin^2\theta -(r^2+a^2)^2\over \Delta\Sigma}\hh
g^{rr}={\Delta\over \Sigma}\, ,
\ee
one gets
\be
-(\nabla W_{\pm})^2={a^2\cos^2\theta\over \Sigma}\ge 0\, .
\ee
For a rotating black hole the equality is valid only when $\theta=\pi/2$. In other words, the gradient $\nabla W_{\pm}$ is always non-spacelike. It is timelike everywhere outside the equatorial plane and is null on this plane. We call the coordinates $W_{-}$ and $W_{+}$ constructed by using the radial star coordinate $r_{\star}$ {\em advanced and retarded star time}, respectively.

Suppose $x^{\mu}(\lambda)$ is a null curve in the Kerr spacetime, then the change of the coordinate $W_{\pm}$ along this curve is
\be
\frac{dW_{\pm}}{d\lambda}=\frac{\partial W_{\pm}}{\partial x^{\alpha}}\frac{dx^{\alpha}}{d\lambda}\geq0\, .
\ee
The equality is valid only if a null curve crosses the equatorial plane, where $W_{\pm;\mu}$ is null, and at that point its tangent vector $dx^{\mu}/d\lambda$ is colinear to
\be
W_{\pm}^{;\mu}|_{\theta=\pi/2}=-(r^2\Delta)^{-1}(H^2,\pm H\Delta,0,2aMr)\, .
\ee
If one excludes these special null curves, one can use both the star time coordinates $W_{+}$ and $W_{-}$ as monotonically increasing parameters along null curves.
The Kerr metric \eq{k2} in the coordinates $x^{\alpha}=(W_{\pm},r,\theta,\phi)$ remains the same with the only change
\ba
&&dt-g_{i}dx^i\to dW_{\pm}-\hat{g}_{i}dx^i\, ,\\
&&\hat{g}_{i}=\mp\frac{H}{\Delta}\delta_{i}^{~r}-\frac{2aMr}{\Sigma h}\sin^{2}\theta\delta_{i}^{~\phi}\,.
\ea

\subsection{Dimensionless form of the equations}

The mass parameter $M$ in the Kerr metric determines the scale. In what follows, it is convenient to use the dimensionless form of the dynamical equations given above. Instead of the coordinates $(t,r,\theta,\phi)$ and $W_{\pm}$ we shall use the coordinates $(\tau,x,y,\phi)$ and $w_{\pm}$ defined as follows:
\ba
\tau&\equiv&\frac{t}{M}\hhh x\equiv\frac{M}{r}\hhh y\equiv\cos\theta\,,\nn
w_{\pm}&\equiv&{W_{\pm}\over M}=\tau\pm\int\frac{\eta}{x^{2}\delta}dx\,,\n{19}
\ea
where
\be
\eta\equiv(1+2\alpha^{2}x^{3}+\alpha^{2}x^{2})^{1/2}\hhh \delta\equiv1-2x+\alpha^{2}x^{2}\,.
\ee
Let us denote
\be
\alpha\equiv\frac{a}{M}\hh d\ell^2\equiv M^{-2}dl^2\,,
\ee
then the Kerr metric \eq{k2} takes the form
\ba
ds^2&=&M^2 h\,dS^2\, ,\\
dS^2&=&-(d\tau -\gamma_i\,dz^i)^2+d\ell^2\,.\n{uS}
\ea
Here 
\ba\n{3}
h&=&1-\frac{2x}{\sigma}\hhh\gamma_{i}=-\frac{2\alpha x(1-y^{2})}{(\sigma -2x)}\delta_{i}^{~\phi}\,,\nn
\sigma&\equiv&1+\alpha^{2}x^{2}y^{2}\,,
\ea
$z^i$ are spatial coordinates, $z^i=(x,y,\phi)$, and
\be
d\ell^{2}=\gamma_{xx}dx^{2}+\gamma_{yy}dy^{2}+\gamma_{\phi\phi} d\phi^{2}\, ,\n{4}
\ee
where
\ba
&&\hspace{-0.9cm}\gamma_{xx}=\frac{\sigma^{2}}{x^{4}\delta(\sigma-2x)}\hhh \gamma_{yy}=\frac{\sigma^{2}}{x^{2}(\sigma-2x)(1-y^{2})}\,,\\
&&\hspace{-0.9cm}\gamma_{\phi\phi}=\frac{\delta\sigma^{2}(1-y^{2})}{x^{2}(\sigma-2x)^{2}}\hhh \gamma=\mbox{det}(\gamma_{ij})=\frac{\sigma^{6}}{x^{8}(\sigma-2x)^{4}}\,.
\n{5}
\ea
In these coordinates the horizon $x=x_{H}$ and the ergosphere $x=x_{erg}$ of a Kerr black hole are determined by the relations
\be\n{horerg}
x_H={1\over 1+\sqrt{1-\alpha^2}}\hh
x_{erg}=\frac{1}{1+\sqrt{1-\alpha^2y^2}}\, .
\ee

The equation for propagation of the circularly polarized photons in the ultrastationary spacetime \eq{uS}, written in the $(3+1)$ form and expressed in the dimensional coordinates has the following form [cf. Eqs. \eq{B2}, \eq{B3}, and \eq{fff}]:
\ba
{D^2{\BM{z}}\over d\ell^2}&=&\left[\frac{d{\BM{z}}}{d\ell}\times \BM{f}_{\ve}\right]\,,\n{18a}\\
\frac{d\tau}{d\ell}&=&1+\left(\BM{\gamma},\frac{d\BM{z}}{d\ell}\right)\,,\n{18b}\\
\BM{f_{\ve}}&=&\mbox{curl}\,\BM{\gamma}+\ve\mbox{curl}\,\mbox{curl}\,\BM{\gamma}\hh
\ve\equiv\pm \frac{1}{2\omega M}\,.\n{18c}
\ea
As we explained above, $\omega$ is the frequency of a photon at infinity, and the sign ($\pm$) defines its helicity, which corresponds to the right ($+$) and to the left ($-$) circularly polarization of an electromagnetic wave. We always assume that $|\ve|\ll1$. 
In the Kerr metric the non-vanishing components of $\mbox{curl}\,\BM{\gamma}$ and $\mbox{curl}\,\mbox{curl}\,\BM{\gamma}$ are given by
\ba
(\mbox{curl}\,\BM{\gamma})_{x}&=&\frac{4\alpha xy}{\sigma(\sigma-2x)}\, ,\\ (\mbox{curl}\,\BM{\gamma})_{y}&=&\frac{2\alpha x^{2}(1-\alpha^{2}x^{2}y^{2})}{\sigma(\sigma-2x)}\,,\\
(\mbox{curl}\,\mbox{curl}\,\BM{\gamma})_{\phi}&=&\frac{4\alpha x^{4}\delta(1-y^{2})}{\sigma(\sigma-2x)^{2}}\,.\n{7}
\ea

For $\ve=0$, that is, for null geodesics of the Kerr metric, Eqs. \eq{18a}~--~\eq{18c} can be written in the first order form
\ba
\frac{dx}{d\ell}&=&\sigma_{x}\frac{x^{2}}{\sigma^{2}}(\sigma-2x)\sqrt{{\cal X}}\hhh \sigma_{x}=\pm1\,,\n{12a}\\
\frac{dy}{d\ell}&=&\sigma_{y}\frac{x^{2}}{\sigma^{2}}(\sigma-2x)\sqrt{{\cal Y}}\hhh \sigma_{y}=\pm1\,,\n{12b}\\
\frac{d\phi}{d\ell}&=&\frac{(p-g_{\phi})}{\gamma_{\phi\phi}}\,,\n{12c}\\
\frac{d\tau}{d\ell}&=&1+\frac{g_{\phi}(p-g_{\phi})}{\gamma_{\phi\phi}}\,.\n{12d}
\ea
Here
\ba
{\cal X}&=&[1-\alpha x^{2}(p-\alpha)]^{2}-x^{2}\delta[q+(p-\alpha)^{2}]\,,\nn
{\cal Y}&=&(q+\alpha^{2}y^{2})(1-y^{2})-p^{2}y^{2}\,,\n{12e}
\ea
where $p$ and $q$ are dimensionless conserved quantities,
\be
p\equiv\frac{L}{EM}\hhh q\equiv\frac{Q}{E^{2}M^{2}}\,.\n{14}
\ee

\section{Scattering problem in the Kerr spacetime}

\subsection{Scattering data}

Our goal is to study propagation of a circularly polarized photon in the Kerr spacetime. Its trajectory is a solution of Eqs. \eq{18a}-\eq{18c}. We always assume that the coefficient $\ve$ which controls the relative strength of the polarization corrections is small, and in the dynamical equations consider the terms with non-vanishing $\ve$ as a perturbation. Let us focus on the scattering problem. We assume that a photon begins its motion at the past null infinity ${\cal J}^{-}$, propagates towards a Kerr black hole, gets scattered, and finally escapes to the future null infinity ${\cal J}^{+}$. Certainly, some of the incoming photons can be captured by the black hole and do not reach ${\cal J}^{+}$. We do not consider them in this paper. Moreover, we use the $(3+1)-$ formalism (see, \cite{FS}), which is well defined in the domain where the Killing vector is timelike. Hence, we assume that the photons propagate outside the black hole ergosphere.

The spin-optical interaction described by the effective force $\BM{f_{\ve}}$ is proportional to $\mbox{curl}\,\mbox{curl}\,\BM{g}$. It grows when a photon approaches the black hole and reaches its maximum near its radial turning point. Far away from the Kerr black hole this extra force rapidly falls down as
\be
|\mbox{curl}\,\mbox{curl}\,\BM{g}|\sim\frac{4aM^{2}}{r^{5}}\sin\theta\,.
\ee
Hence, the polarization dependent corrections to a photon trajectory become negligible in the asymptotic domain where $r\to\infty$. In other words, photons of the left and right helicity practically follow null geodesics in the asymptotic region, and they can be uniquely specified by the same `scattering data' as unpolarized photons.

Let us remind these `scattering data'. To be concrete, let us consider the future null infinity ${\cal J}^{+}$. This is a three-dimensional surface which can be parameterized by the angular coordinates $\theta$ and $\phi$, which specify the direction of motion of an outgoing photon and the moment of its arrival to ${\cal J}^{+}$, $W_+=M w_+$.
It should be emphasized that the star time coordinates $w_{\pm}$ are defined up to a constant. To fix this constant, one needs to chose the initial radial coordinate in the integral for $r_{\star}$ in \eq{w}. We shall discuss this point in Sec. V. In fact, a set of parallel moving photons arrive at the same point of ${\cal J}^{+}$. The `members' of this set can be parametrized by two additional parameters, $\xi$ and $\zeta$, called the impact parameters.

\begin{figure}[htb]
\begin{center}
\includegraphics[width=6.0cm]{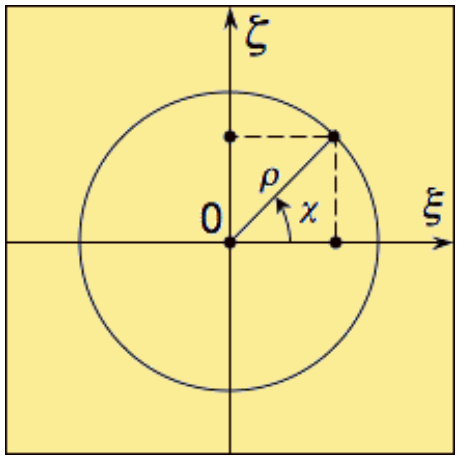}
\caption{Impact plane. The direction of $\xi$ is in the positive direction of $\phi$ (direction of the black hole rotation), and the direction of $\zeta$ is in the positive direction of $\theta$. A circle of $\rho=const$ corresponds to a constant asymptotic value of a photon total angular momentum.}\label{F3}
\end{center}
\end{figure}

At the large distance  $r_o$ solutions of the photon equation of motion in the Kerr metric
are of the form [cf. Eqs. \eq{11a}--\eq{11d}]
\ba
{dr\over dt}&\sim&E\,,\\
{d\phi\over dt}&\sim & {L\over E r_o^2 \sin^2\theta_o}\, ,\\
{d\theta\over dt}&\sim &\pm{1\over E r_o^2}\sqrt{Q +E^2 a^2\cos^2\theta_o-L^2\cot^2\theta_o}\, .
\ea
The angles of displacement of the photon in the $\phi$ and
$\theta$ directions are $~r_o\sin\theta_o\,{d\phi/dt}$ and
$r_o\,{d\theta/dt}$, respectively. These angles decrease as
$r_o^{-1}$. Multiplying these quantities by $r_o$ one obtains the impact parameters $(M\xi,M\zeta)$. The dimensionless impact parameters $(\xi,\zeta)$ are related to the asymptotic integrals of motion $(p,q)$ as follows:
\be\n{eqx}
\xi_{o}=-{p\over\sin\theta_o}\hhh\zeta_{o}=\sigma_{\theta}\sqrt{q +\alpha^2\cos^2\theta_o-p^2 \cot^2\theta_o}\,.
\ee
The impact vectors $\BM{\rho}=(\xi,\zeta)$ form a two-dimensional impact plane shown in Fig.~\ref{F3}.
In what follows, it is convenient to use the polar coordinates $(\rho,\chi)$ on this plane.
To summarize, a photon arriving at ${\cal J}^{+}$ is uniquely specified by the five dimensionless parameters ${\cal D}_o=\{w_{-o},y_{o}=\cos\theta_{o},\phi_{o},\xi_{o},\zeta_{o}\}$. We call these parameters the scattering data for an outgoing photon. A similar set
${\cal D}_e=\{w_{+e},y_{e}=\cos\theta_{e},\phi_{e},\rho_{e},\chi_{e}\}$ forms the scattering data for an incoming photon.

The scattering problem for photons in the Kerr spacetime can be formulated as follows. Determine
${\cal D}_e$ for a photon which begins its motion with the parameters ${\cal D}_o$. One may consider a map $\Psi$,
\be
\Psi: {\cal D}_e\to {\cal D}_o\,,
\ee
as a classical analogue of an $S-$matrix. In the absence of spin-optical effects (for null geodesics) the scattering problem can be reduced to elliptic integrals. In the presence of spin-optical corrections, when the equations for a null ray contain the parameter $\ve$, this map is slightly modified and becomes $\Psi(\ve)$. We would like to study this modified spin-dependent map as a function of the parameter $\ve$. In accordance with our main assumption we restrict ourselves to  effects linear in $\ve$. Note that in the expansion of $\Psi(\ve)$ with respect to $\ve$, the zero and the first order in $\ve$ terms are nontrivial functions of the coordinates $(w_{\pm}, x, y, \phi)$, which are defined by a solution to the dynamical equations \eq{18a}--\eq{18c}.  
\begin{figure}[htb]
\begin{center}
\includegraphics[width=6.5cm]{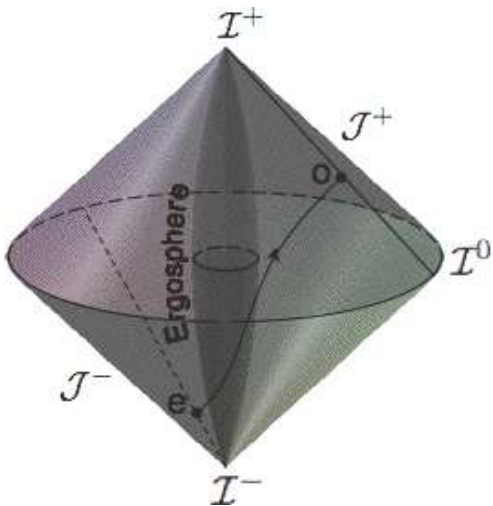}
\caption{Diagram illustrating scattering of photons by a Kerr black hole. The angular coordinate $\theta$ is suppressed. A photon trajectory is shown by the curve connecting the points $e$ (emitter) at ${\cal J}^{-}$ and $o$ (observer) at ${\cal J}^{+}$.}\label{F1}
\end{center}
\end{figure}

The scattering of photons by a Kerr black hole is schematically illustrated in Fig. \ref{F1}.
If the black hole is absent, i.e., if the spacetime is Minkowski one, then a photon trajectory from the emitter to the observer is a straight line. In this case, one can easily define the final angular coordinates $(\phi_{o'},\theta_{o'})$ at ${\cal J}^{+}$ in terms of the initial ones $(\phi_{e},\theta_{e})$ at ${\cal J}^{-}$ as follows:
\be
\theta_{o'}=\pi-\theta_{e}\hhh \phi_{o'}=\pi+\phi_{e}\,.
\ee
In the presence of a black hole, a photon trajectory will deviate from the straight line, so that $\phi_{o}-\phi_{e}>\pi$. To measure such a deviation, one can define the bending angles $\theta_{o}-(\pi-\theta_{e})$ and $\phi_{o}-(\pi+\phi_{e})>0$. In the absence of spin-optical interaction, photons propagate along null geodesics of the Kerr spacetime. In this case, the bending angles can be calculated analytically. In the case when the angle $\theta$ has only one minimum or maximum, the bending angles were calculated in \cite{ITT} up to the third-order terms $(M/r_{min})^{3}$, $(a/r_{min})^{3}$, where $r_{min}$ corresponds to the radial turning point of a null geodesic.

\subsection{Impact parameters and asymptotic integrals of motion}

There exist two different ways how to specify a null curve reaching ${\cal J}^{\pm}$ at the given values $(w_{\pm},y,\phi)$: (i) either to give its impact parameters, or (ii) to fix its asymptotic integrals of motion $p$ and $q$. Let us discuss this issue in detail. Equation \eq{eqx} establishes the relations between these two different sets of data. Near ${\cal J}^{-}$ one can use the Cartesian coordinates $(t,X,Y,Z)$. In these coordinates
an incoming photon of the energy $E$ has the following 4-momentum defined at ${\cal J}^{-}$:
\ba
p^{\alpha}&=&(E,{\BM p})\hhh \BM{p}^{2}=E^{2}\,,\nn
\BM{p}&=&E(-\sin\theta_{e}\cos\phi_{e},-\sin\theta_{e}\sin\phi_{e},-\cos\theta_{e})\,,
\ea
while the spatial dimensionless vector $\BM{\rho}$ corresponding to the impact parameters $(\xi_e,\zeta_e)$ has the coordinates
\ba
\rho_{X}&=&-\xi_{e}\sin\phi_{e}+\zeta_{e}\cos\theta_{e}\cos\phi_{e}\,,\nn
\rho_{Y}&=&\xi_{e}\cos\phi_{e}+\zeta_{e}\cos\theta_{e}\sin\phi_{e}\,,\nn
\rho_{Z}&=&-\zeta_{e}\sin\theta_{e}\,.
\ea
Calculating the photon total angular momentum
\be
\BM{L}=M\left[\BM{\rho}\times\BM{p}\right]\,,
\ee
we obtain
\be\n{Lz}
p=\rho\cos\chi\sin\theta_{e}\hhh \rho=\sqrt{\xi_{e}^{2}+\zeta_{e}^{2}}=|\BM{L}|/(EM)\,.
\ee
It is possible to show (see, e.g., \cite{FZ}) that in the asymptotically flat region  the Carter constant $Q$  and the asymptotic integrals of motion are related as follows:
\be\n{QLP}
\BM{L}^{2}=Q+L_{Z}^{2}+a^{2}p_{Z}^{2}\, .
\ee
Using the relations \eq{Lz} and \eq{QLP} one obtains
\be\n{qqq}
q=\rho^{2}-\rho^{2}\cos^{2}\chi\sin^{2}\theta_{e}-\alpha^{2}\cos^{2}\theta_{e}\,.
\ee
Equations \eq{Lz} and \eq{qqq} are equivalent to Eq. \eq{eqx}. For $\ve=0$ and the given initial point $(w_{-e},  \theta_{e},\phi_{e})$ these integrals of motion uniquely determine a null geodesic. For $\ve\ne 0$, $p$ is still an integral of motion, while $q$ is only asymptotic invariant. However, $(p,q)$ defined at $(w_{-e},\theta_{e},\phi_{e})$ uniquely determine a null ray solution of the perturbed equations \eq{18a}--\eq{18c}.

Let us fix the star time $w_{-e}$ of a photon emission and the spherical coordinates $(\theta_{e},\phi_{e})$ of the emission point. In what follows, we discuss the following three problems:
\begin{enumerate}
\item How does the photon bending angle depend on its polarization?
\item How does position of the image of a photon arriving to an observer at $(\theta_{o},\phi_{o})$ depend on its polarization?
\item How does the arrival time of such photons depend on their polarization?
\end{enumerate}

\section{Polarization shift of the bending angles}

\subsection{Shift parameters}

Let us discuss the first problem. After fixing the parameters $(w_{-e},\theta_{e},\phi_{e})$ there still remain two impact parameters. For $\ve=0$ and the chosen impact parameters $(\xi_{e},\zeta_{e})$  the corresponding null geodesic arrives to ${\cal J}^+$ at  some moment of the retarded star time $w_{+o}$ in the direction $(\theta_{o},\phi_{o})$ with the impact parameters $(\xi_o,\zeta_o)$. We focus our attention on the bending angles of a photon which are determined by $(\theta_{o},\phi_{o})$. For circularly polarized photons ($\ve=\pm|\ve|\ne 0$) with the same initial scattering data the arrival angles $(\theta_{(\pm)},\phi_{(\pm)})$ are slightly different,
\ba\n{bangle}
\theta_{(\pm)}=\theta_{o}\pm|\ve|\Delta\theta_{o}+O(\ve^2)\,,\nn
\phi_{(\pm)}=\phi_{o}\pm|\ve|\Delta\phi_{o}+O(\ve^2)\,.
\ea
The quantities $\Delta\theta_{o}$ and $\Delta\phi_{o}$ allow us to measure polarization shift of the bending angles. We shall analyze how the polarization shift of the bending angles depends on the impact parameters $(\xi_{e},\zeta_{e})$.

Since for $\ve\ne 0$ we do not know analytical solutions, to solve this problem we use numerical methods.
There are certain restrictions which have to be taken into account in the numerical calculations. First of all, we must chose the initial impact parameters so that photons are not captured by the black hole. Note, that in the calculation of the polarization contribution to the equation of motion the ultrastationary form of the metric was used \cite{FS}. For this reason, trajectory of a photon should not enter the ergosphere. Moreover, in the very vicinity of the ergosphere the effective force $f_{\ve}$ is growing fast, and the validity of the adapted linear in $\ve$ approximation can be violated. To control the validity of the approximation  we always check that the expansion \eq{bangle} is valid and the deflection of the numerical result from the linear law is small.

Before we describe the adopted numerical scheme, let us make a couple of additional remarks. We use the spherical coordinates which are evidently behave badly near the axis of symmetry $\theta=0,\pi$. In particular, for null rays passing close to the axis of symmetry, a relatively large value of $\Delta\phi_{o}$ does not mean that the bending angle is large. A better measure for the polarization shift of the bending angles is the quantity
\be
\Delta\alpha\equiv\left[(\Delta\theta_{o})^{2}+\sin^{2}\theta_{o}(\Delta\phi_{o})^{2}\right]^{1/2}\, .
\ee
This is the invariant distance on a unit (celestial) sphere between its two close points, divided by $|\ve|$.
To keep trace of the direction of the displacement we shall use the angle $\beta$ between the displacement vector and the $\phi-$direction,
\be
\tan \beta=\frac{\Delta\theta_{o}}{\sin\theta_{o}\,\Delta\phi_{o}}\,.
\ee

\subsection{Approach and results}

The details of the scheme of the numerical calculations are given in Appendix A. Here we briefly discuss the main important steps of these calculations, the adopted choice of the initial data, and the obtained results.
\begin{figure*}[htb]
\begin{center}
\ba
&&\hspace{0cm}\includegraphics[width=6cm]{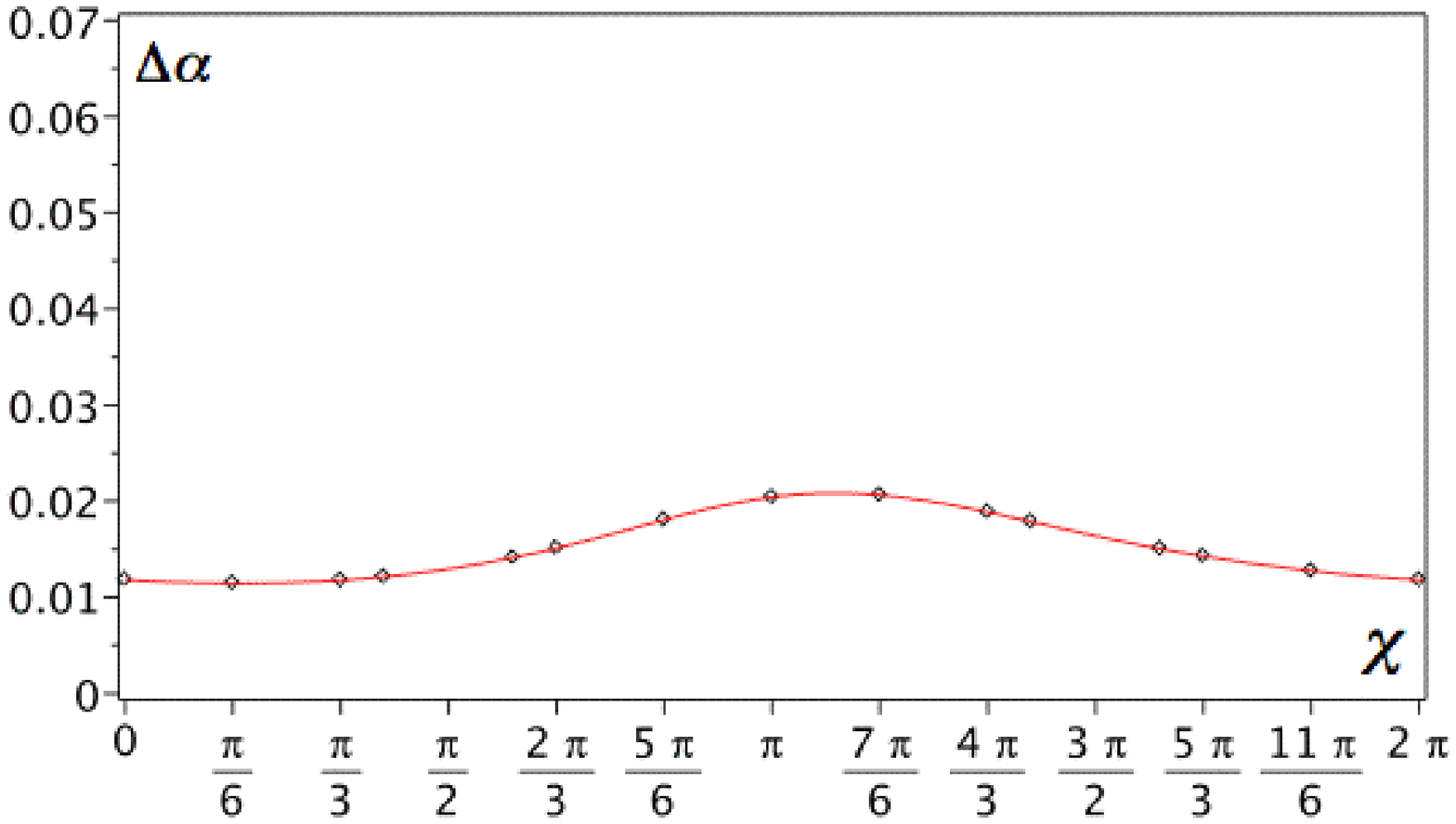}
\hspace{2cm}\includegraphics[width=6cm]{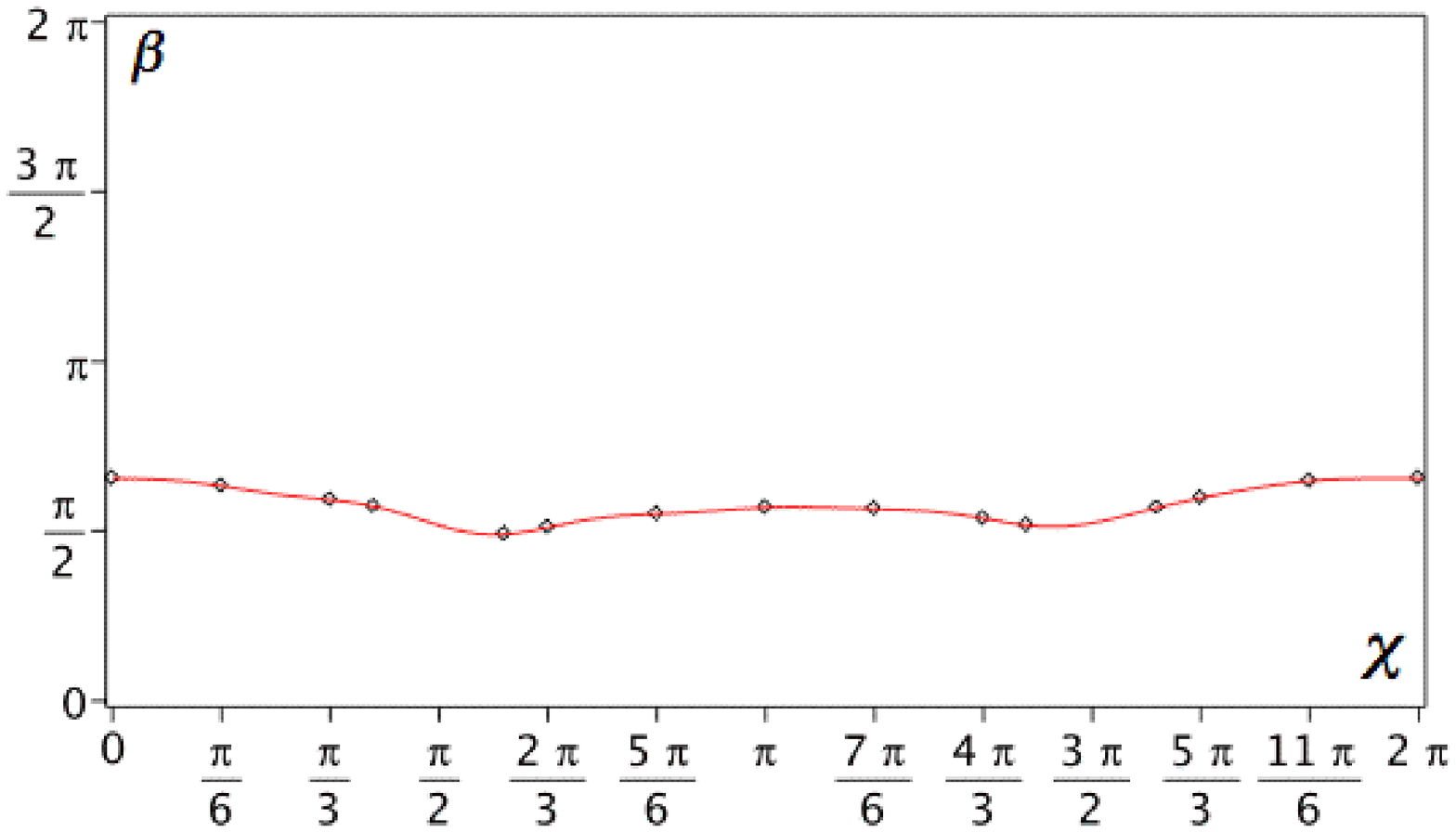}\nn
&&\hspace{2.8cm}({\bf a1})\hspace{7.5cm}({\bf a2})\nn
&&\hspace{0cm}\includegraphics[width=6cm]{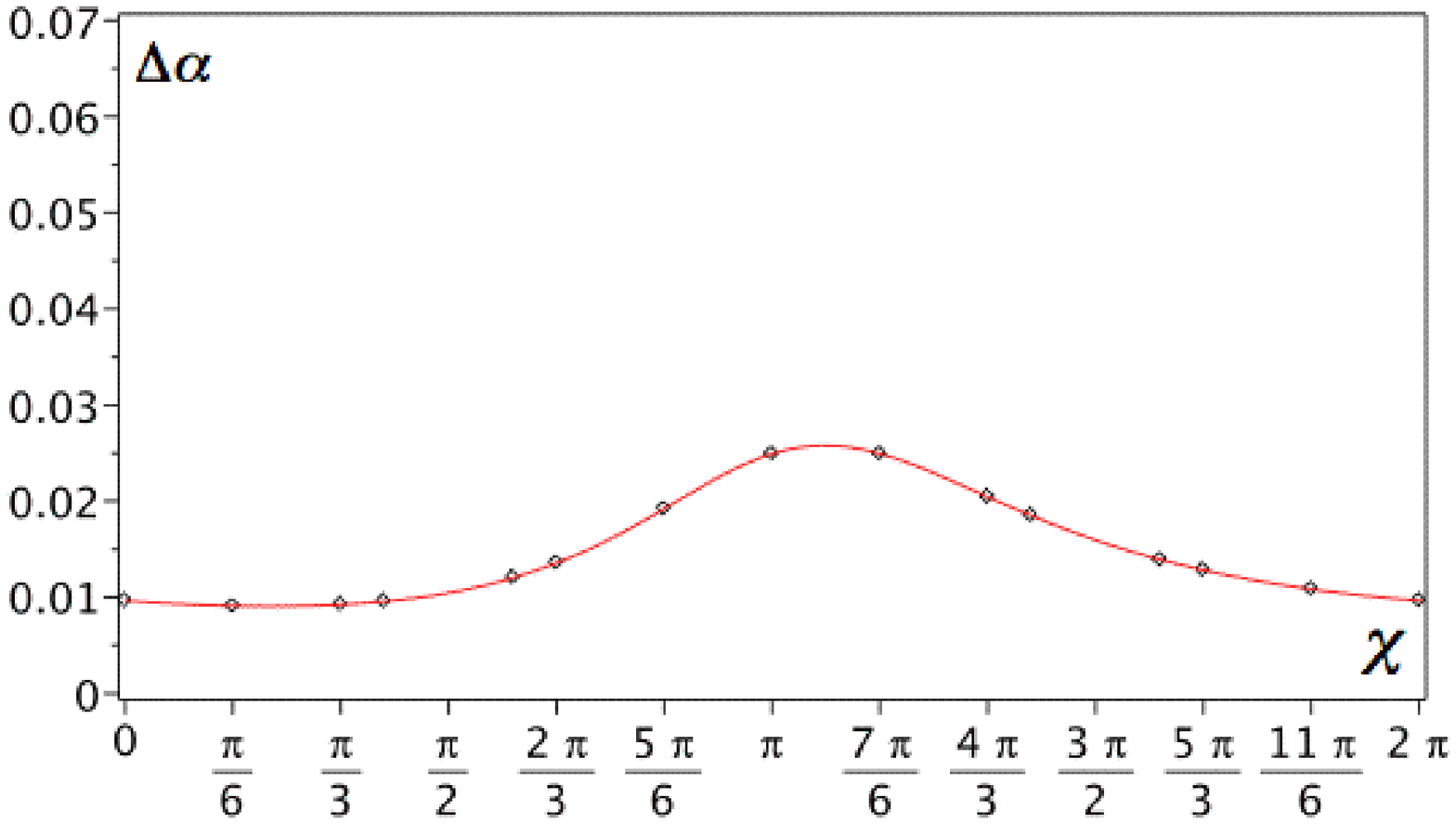}
\hspace{2cm}\includegraphics[width=6cm]{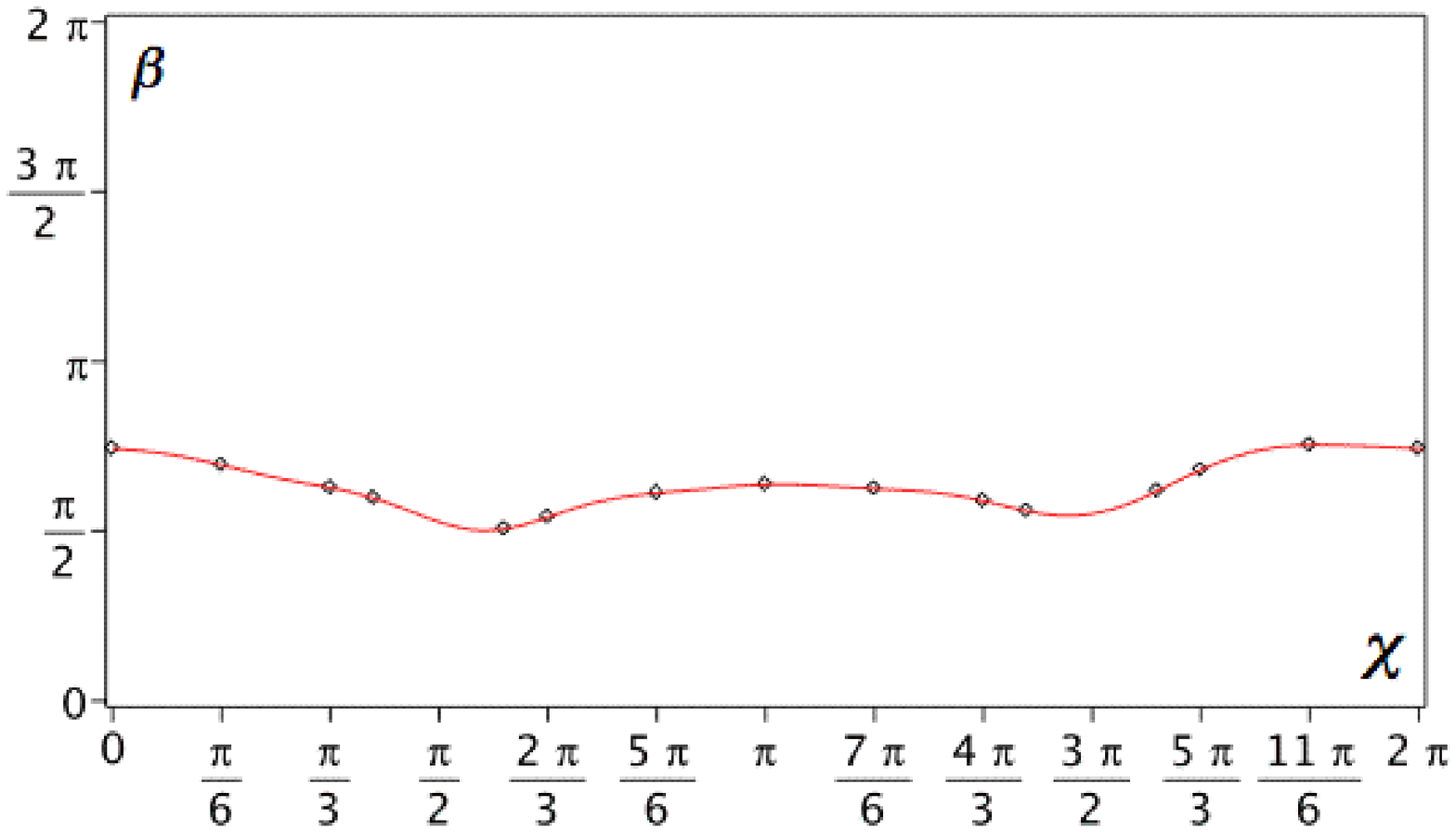}\nn
&&\hspace{2.8cm}({\bf b1})\hspace{7.5cm}({\bf b2})\nn
&&\hspace{0cm}\includegraphics[width=6cm]{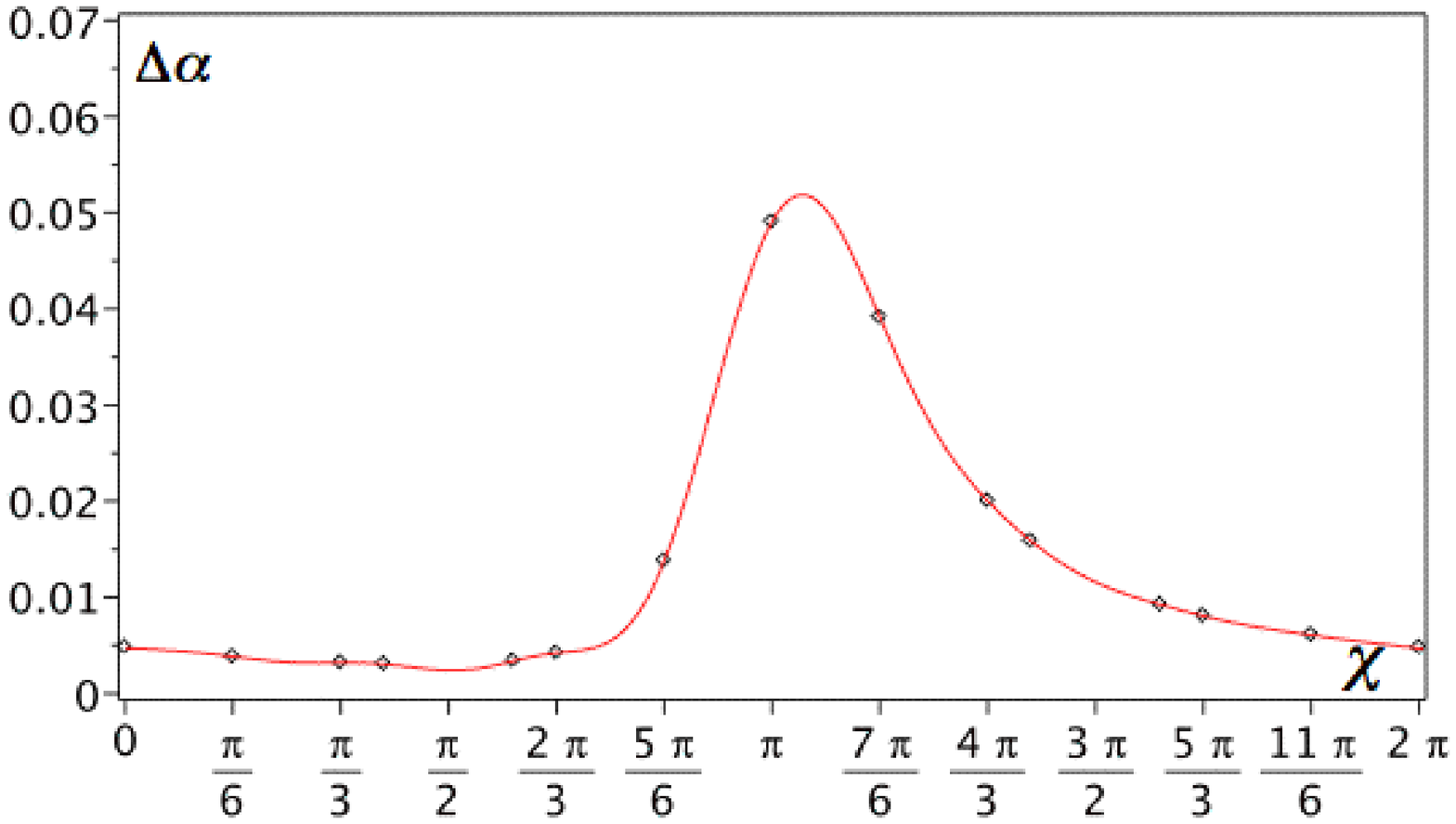}
\hspace{2cm}\includegraphics[width=6cm]{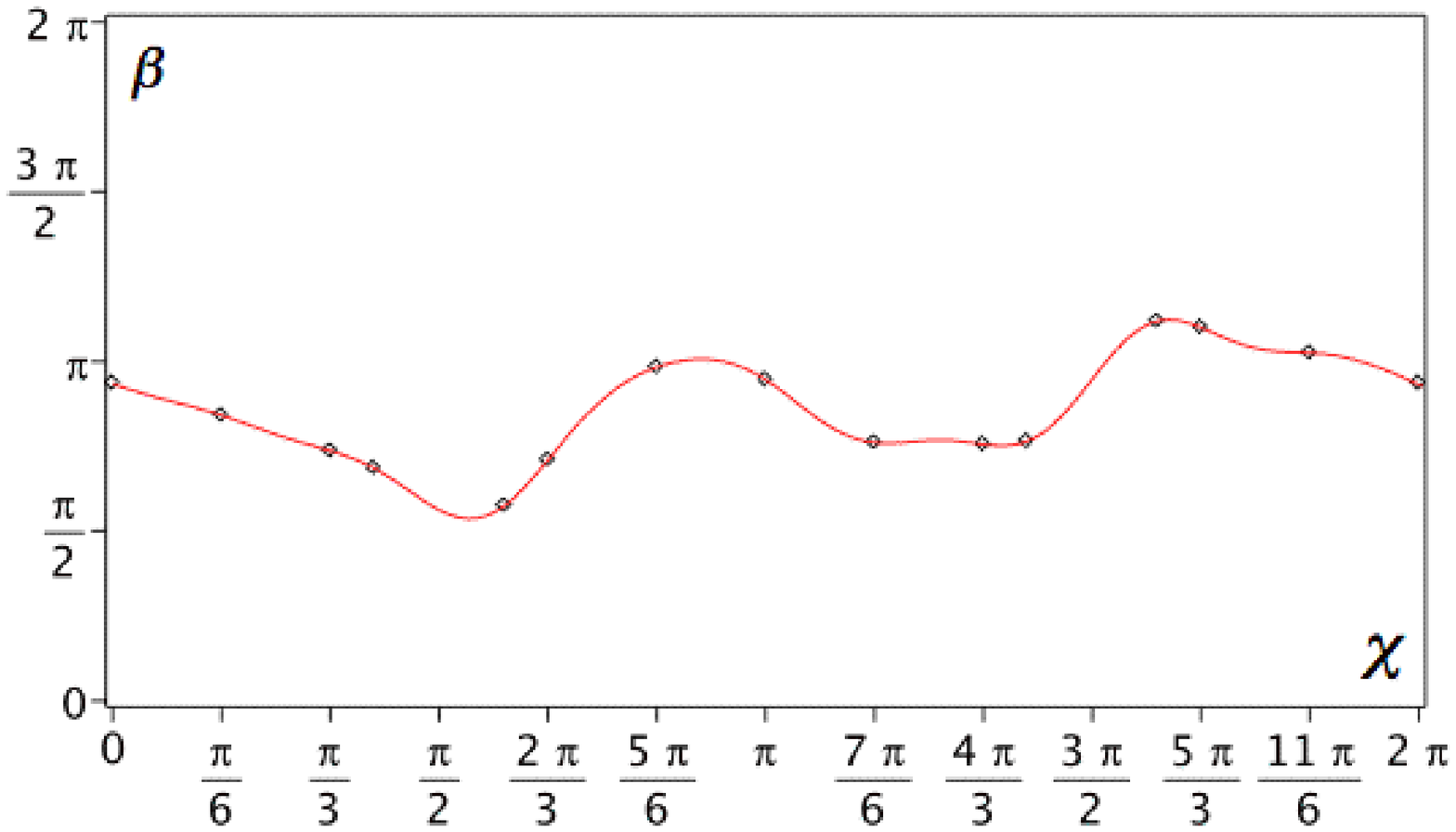}\nn
&&\hspace{2.8cm}({\bf c1})\hspace{7.5cm}({\bf c2})\nn
&&\hspace{0cm}\includegraphics[width=6cm]{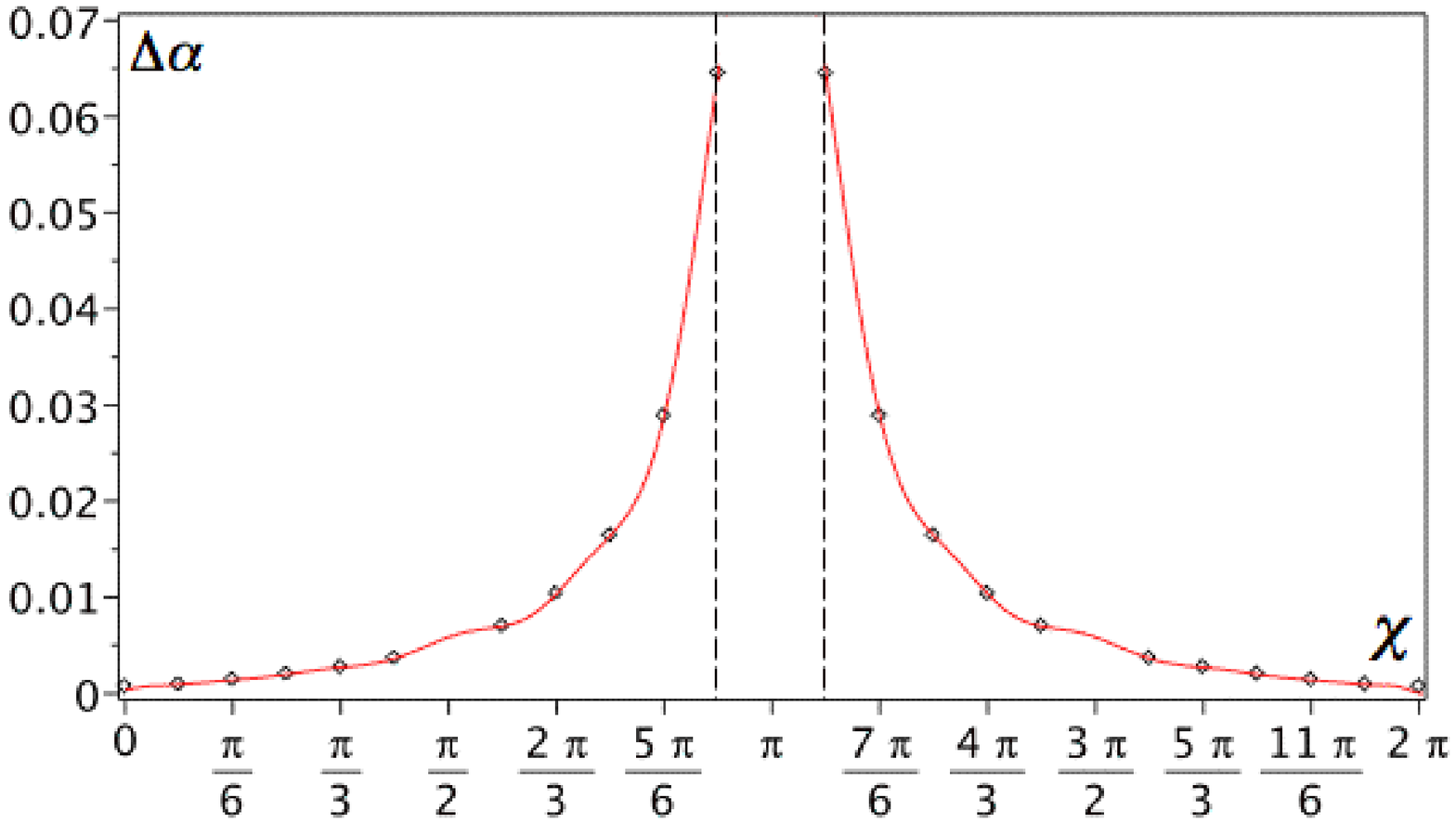}
\hspace{2cm}\includegraphics[width=6cm]{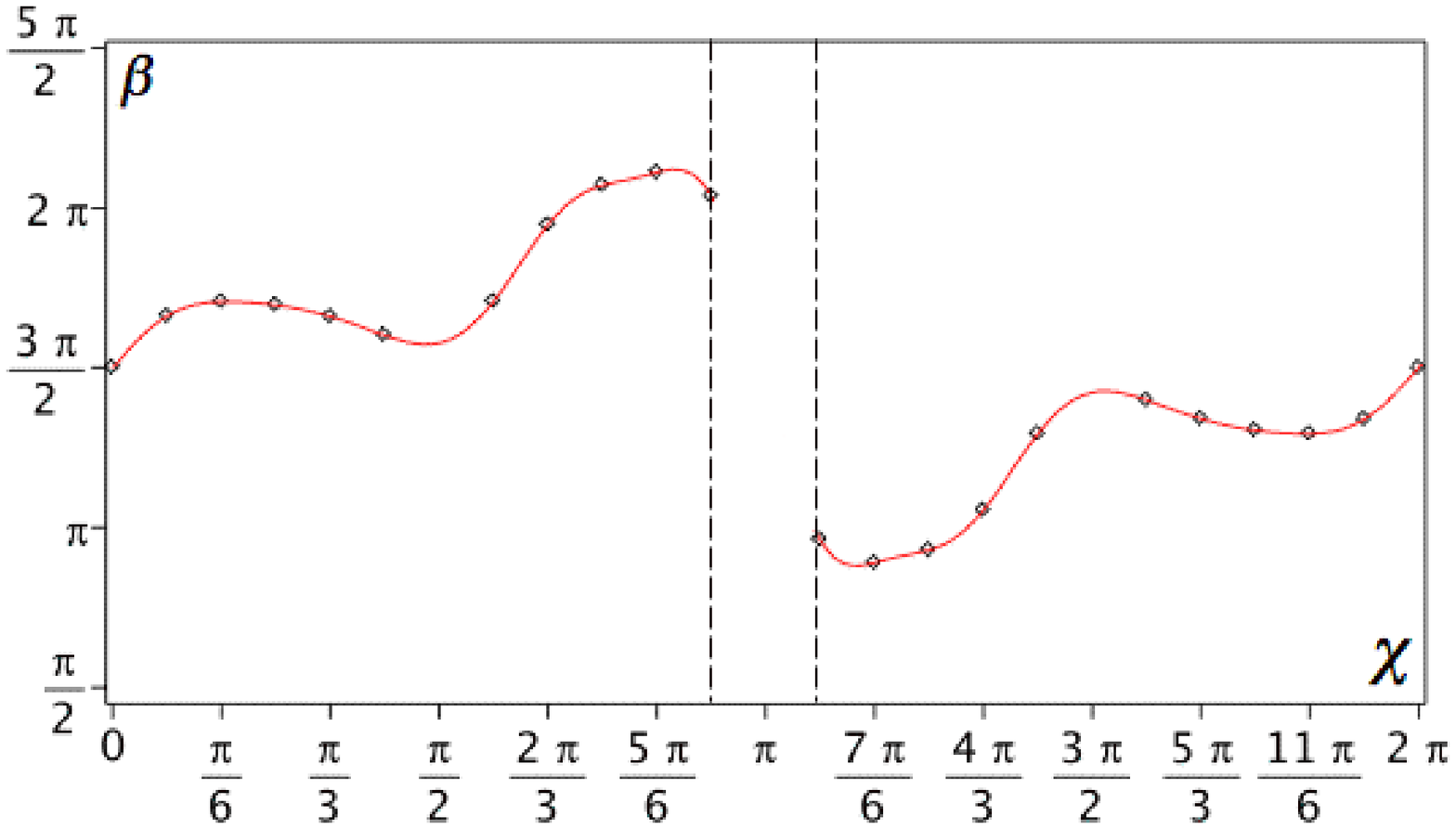}\nn
&&\hspace{2.8cm}({\bf d1})\hspace{7.5cm}({\bf d2})\nonumber
\ea
\caption{Polarization shift of the bending angles. The left column contains plots showing dependence of the shift parameter $\Delta\alpha$ (shift angle divided by $|\ve|$) on the angle $\chi$ for four different inclination angles $\theta_e$. The right column contains similar plots for the parameter $\beta\in[0,2\pi)$, which defines the direction of the shift.  Figures ({\bf a1}) and ({\bf a2}) correspond to $\theta_{e}=\pi/10$, Figs. ({\bf b1}) and ({\bf b2}) correspond to $\theta_{e}=\pi/6$, Figs. ({\bf c1}) and ({\bf c2}) correspond to $\theta_{e}=\pi/3$, and Figs. ({\bf d1}) and ({\bf d2}) correspond to $\theta_{e}=\pi/2$. The vertical dashed lines in Figs. ({\bf d1}) and ({\bf d2}) cut out the values of $\chi$ close to $\pi$.}\label{F4}
\end{center}
\end{figure*}
\begin{figure}[htb]
\begin{center}
\includegraphics[width=6.0cm]{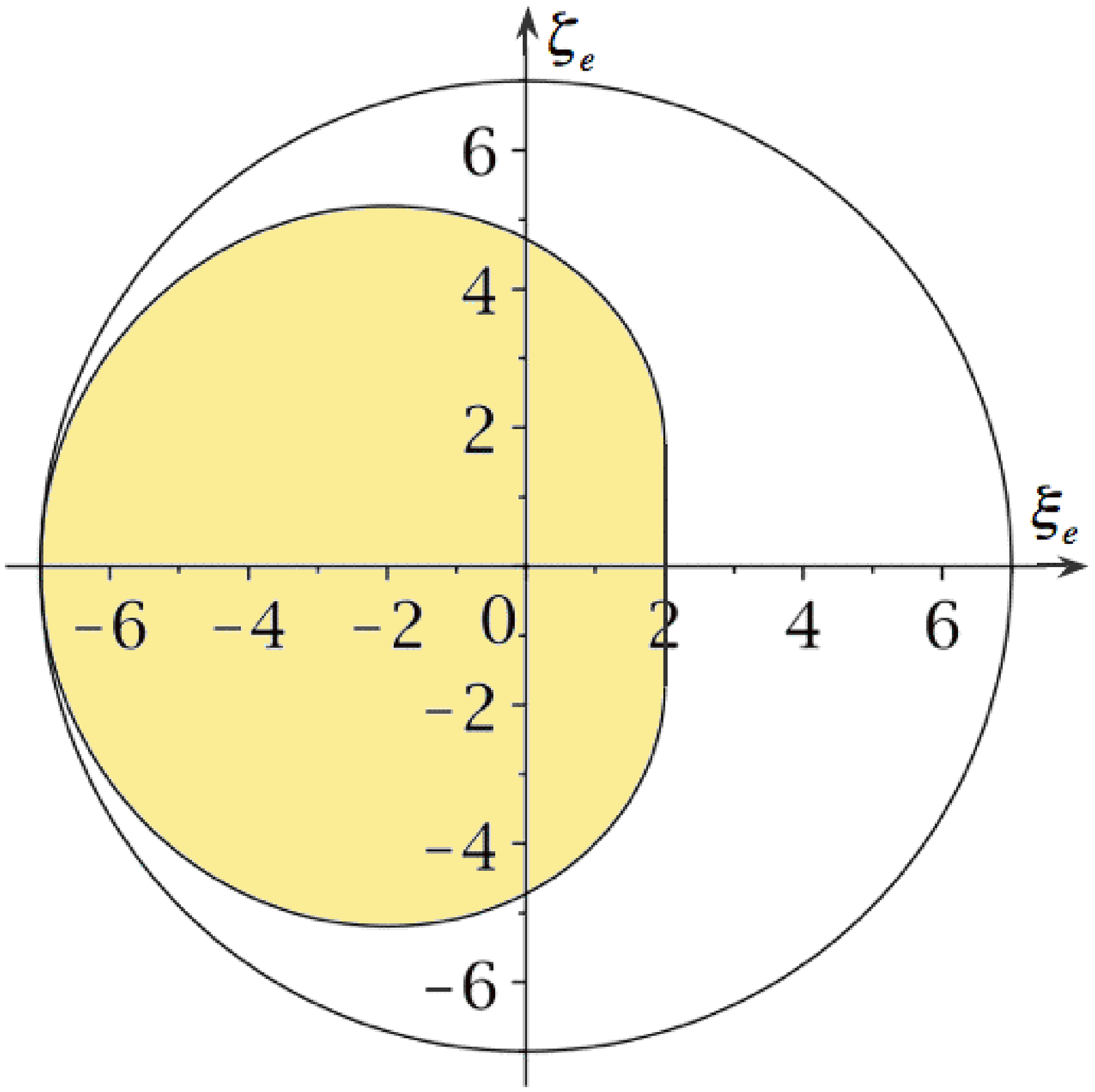}
\caption{Capture domain for an extremal black hole ($\alpha=1$) and the inclination angle $\theta_e=\pi/2$. The circle defines the `tube' of the null curves with the impact radius $\rho=7$. Incoming null geodesics with impact parameters within the shadowed region (capture domain) are captured by the black hole, while for the impact parameters outside this region scattered photons reach ${\cal J}^+$.}\label{Fs}
\end{center}
\end{figure}
We start  the numerical integration by choosing the initial data for a null ray, that is, by fixing
${\cal D}_e=\{w_{-e},y_{e}=\cos\theta_{e},\phi_{e},\rho_{e},\chi_{e}\}$ of an incoming at ${\cal J}^-$ photon. Using the axial symmetry of the problem we put $\phi_e=0$. The Kerr metric is invariant under the reflection with respect to the equatorial plane $y\to -y$. Equations \eq{18a}--\eq{18c} also respect this symmetry, if one simultaneously changes $\ve\to -\ve$ in these equations. For this reason, we consider only the values of $\theta_{e}$ lying in the upper semisphere $0<\theta_{e}\le \pi/2$. To avoid the problem with bad behavior of the spherical coordinate $\phi$ near the symmetry axis, we always take a non-vanishing inclination angle $\theta_{e}$.

For the incoming null ray we use the advanced star time $w_{-}$ as a parameter along the ray.
The inverse radius $x$ increases from its initial value $x=0$ at ${\cal J}^-$ to some  value
$x_{max}=M/r_{min}$ corresponding to the radial turning point, where the radial coordinate reaches its minimal value $r_{min}$. After the turning point, we use the retarded star time $w_{+}$ as a parameter along the ray (for details see Appendix A). In order to make $w_{\pm}$ well defined we chose the lower value in the integral in \eq{w}, which determines $r_{\star}$, to be equal to $2M$. Thus, at the turning point $r=r_{min}$ we have
\be\n{wpm}
w_{+}|_{r_{min}}=(w_{-}-2r_{\star}/M)|_{r_{min}}\,.
\ee
At the end point at ${\cal J}^+$, where $x$ again takes zero value, the retarded star time $w_{+}$ determines the time of arrival of the null ray.
In the coordinates \eq{19} one has
\be\n{time}
w_{+}|_{{\cal J}^{+}}=w_{-}|_{{\cal J}^{-}}+\Delta w_{-}+\Delta w_{+}-2\int^{1/2}_{x_{max}}\frac{\eta}{x^{2}\delta}dx\,.\ee
Here $\Delta w_{-}$ and $\Delta w_{+}$ are calculated along the ingoing and outgoing null rays, respectively.

The numerical procedure described in Appendix A allows one to calculate the polarization shift of the benging angles. Here we present only the results for the extremal Kerr spacetime ($a=M, \alpha=1$), where the expected effects are the most profound. To illustrate the dependence of the shift on the inclination angle $\theta_e$, we present the results for its four different values, $\theta_e=\pi/10,  \pi/6, \pi/3$, and $\pi/2$.

For a chosen inclination angle $\theta_e$ one still needs to specify the impact parameters, which belong to the two-dimensional impact plane. In order to reduce the space of parameters we use the following approach. We fix the impact radius $\rho$ and perform calculations for a discrete set of the angles $\chi$. By fixing $\rho$ we keep  the value of the asymptotic total angular momentum $|\BM{L}|$ constant [see Eq. \eq{Lz}]. Figure~\ref{F4} shows the results obtained for a special choice of $\rho=7$. The argument in the plots presented in the Figure is the angular parameter $\chi$, which parametrises the `tube' of null curves with fixed $\rho$. To better illustrate our results, we connected the points generated by the numerical computation into the smooth periodic in $\chi\in[0, 2\pi)$ curves constructed by the spline interpolation of the ninth degree and followed by the polynomial interpolation using the Lagrange method. The function $\beta$ is a continuous function of $\chi$ which takes its values on a circle $S^{1}$.  To avoid its discontinuity in the special case shown in Fig. ({\bf d2}) for the inclination angle $\theta_{e}=\pi/2$, we shifted its range from $[0, 2\pi)$ to $[\pi/2, 5\pi/2)$.

For a chosen inclination angle $\theta_e$ a photon trajectory is defined by the impact parameters $(\xi_e,\zeta_e)$. For small impact parameters the photon is captured by the black hole. This means that one can identify two regions on the impact plane with qualitatively different photon trajectories. We call the first region, for which the capture occurs, {\em a capture domain}. It is easy to show that by the reflection $t\to -t,  \phi\to -\phi$ this region is connected with a region on the outgoing impact plane, which is called black hole shadow (see, e.g., \cite{Bardeen}, \cite{HM}, and \cite{FZ}).
The capture domain for the inclination angle $\theta_e=\pi/2$ is shown in Figure~\ref{Fs}. This is the case when the incoming photons begin their motion parallel to the equatorial plane with $\phi_e=0$. The circle in this picture defines the `tube' of the null curves with the impact radius $\rho=7$. It touches the capture domain at the point where $\chi=\pi$. Null geodesics emitted with the impact parameters close to this point revolve around the black hole many times before they either get captured or reach a remote observer. This means that for the chosen values of $\rho=7$ and $\chi$ close to $\pi$ the bending angles become large. In plots ({\bf d1}) and ({\bf d2}) we cut out the region of $\chi$ near $\pi$, where the bending angles become larger than $\pi$. One can see that the corresponding polarization shift $\Delta\alpha$ grows fast near $\chi=\pi$ (see plot ({\bf d1}) in Figure~\ref{F4}). However, our calculations show that inside this region $\Delta\alpha$ can reach the value of $\approx 5.6$ still preserving the linearity in $\ve$ with the accuracy $\approx97.5\%$. 

It is possible to show that for other values of the inclination angle $\theta_{e}$ the impact radius $\rho$ corresponding to the points inside the corresponding capture regions is always smaller than $7$. For the data presented in Figure~\ref{F4} we kept the same value $\rho=7$ and changed the inclination angle. All the plots are periodic in the impact angle $\chi$ with the period $2\pi$. When the inclination angle changes from $\pi/2$ to 0 the plots for $\Delta\alpha$ become more smooth and the maximal value of $\Delta\alpha$ decreases.
For $\chi=\pi/2$ and $3\pi/2$ the azimuthal angular momentum vanishes, i.e., $p=0$. Null geodesics with such value of $p$ pass through the axis of symmetry [cf. Eqs. \eq{12b} and \eq{12e}]. To avoid the problem of the coordinate singularity of $\phi$, we performed the calculations only in the vicinity of these points.

Plots shown on the right side of Figure~\ref{F4} illustrate the dependence of the angle $\beta$ on the impact angle $\chi$. One can see that the range of $\beta$ grows when one considers inclination angles close to $\pi/2$ (equatorial plane).

As it was mentioned in the Introduction, Mashhoon estimated the polarization shift of the bending angles in the weak field approximation \cite{,Mashhoon_93,Ma:74a,Ma:75}.  In our notations, the estimated relation (see Eq. \eq{sepang}) can be written in the following form:
\be
\mbox{separation angle}\sim 2\Delta\alpha|\ve| \sim \kappa\frac{4\pi\alpha|\ve|}{\rho^{3}}\,,
\ee
where $\kappa$ is a dimensionless parameter. For an extremely rotating black hole ($\alpha=1$) the parameter $\kappa$ can be defined as
\be
\kappa=\frac{\Delta\alpha\rho^3}{2\pi}\,.
\ee
For the chosen value $\rho=7$ one has $\kappa\approx55\Delta\alpha$. According to the data presented in Figure~\ref{F4},  $\Delta\alpha$ changes approximately in the range 0.01~---~0.06, so that the corresponding value of $\kappa$ is roughly between $0.6$ and $3.3$. The increase of the parameter $\kappa$ can be explained by the fact that the rays with larger value of $\Delta\alpha$ come closer to the black hole.

\section{Polarization splitting of images and arrival time}

\subsection{Image of a point-like source}

Let us now discuss a different problem. Let us again consider a spacetime with a rotating black hole. Suppose now that there is a bright point-like object emitting light isotopically and an observer which registers that light. For example, the observer takes a picture of this object by using a telescope. If the light passes close to the black hole, position of the image will depend on impact parameters of the corresponding ray. Suppose now that the emitted light is circularly polarized. Since trajectories of circularly polarized photons are slightly different than those of photons following null geodesics, they arrive to the observer at slightly different angles. As a result, the position of the corresponding images in the picture will be slightly different. 

Let us assume that the distances from the black hole to the emitter and to the observer are large. In this case, in order to describe the polarization dependent splitting of the images it is convenient to use the impact parameters $(\xi_{o},\zeta_{o})$ which determine the position of the image.
In this idealized set-up the emitter is located at ${\cal J}^-$. As before, we put $\phi_e=0$, so that its position is characterized only by the inclination angle $\theta_e$.  Denote the angles which determine direction to the observer by $(\theta_o,\phi_o)$. In general, there exist many null geodesics connecting the emitter and the observer. They differ by a winding number describing how many turns around the black hole a photon makes before reaching the observer. Brightness of the photon images decreases  with increase of the winding number (see, e.g., \cite{FZ}). We shall focus on one of these null geodesics. In this case, the initial and the final impact parameters are uniquely defined. These parameters determine the asymptotic integrals of motion $(p_e,q_e)$ and $(p_o,q_o)$ [see Eqs. \eq{Lz} and \eq{qqq}]. For a null geodesic of the Kerr spacetime these quantities are integrals of motion, so that $p_o=p_e$ and $q_o=q_e$.
\begin{figure*}[htb]
\begin{center}
\ba
&&\hspace{0cm}\includegraphics[width=6cm]{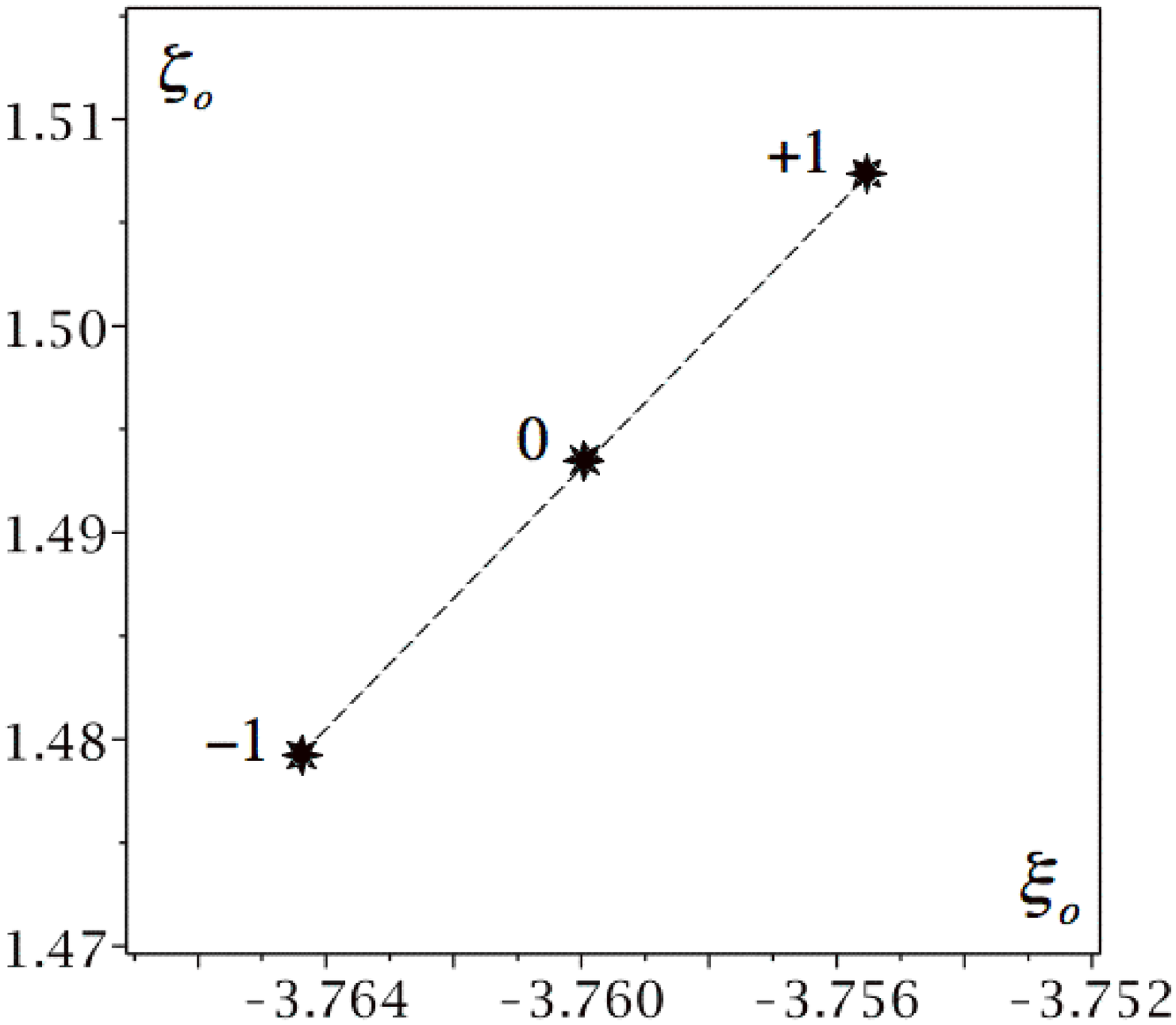}
\hspace{2cm}\includegraphics[width=6cm]{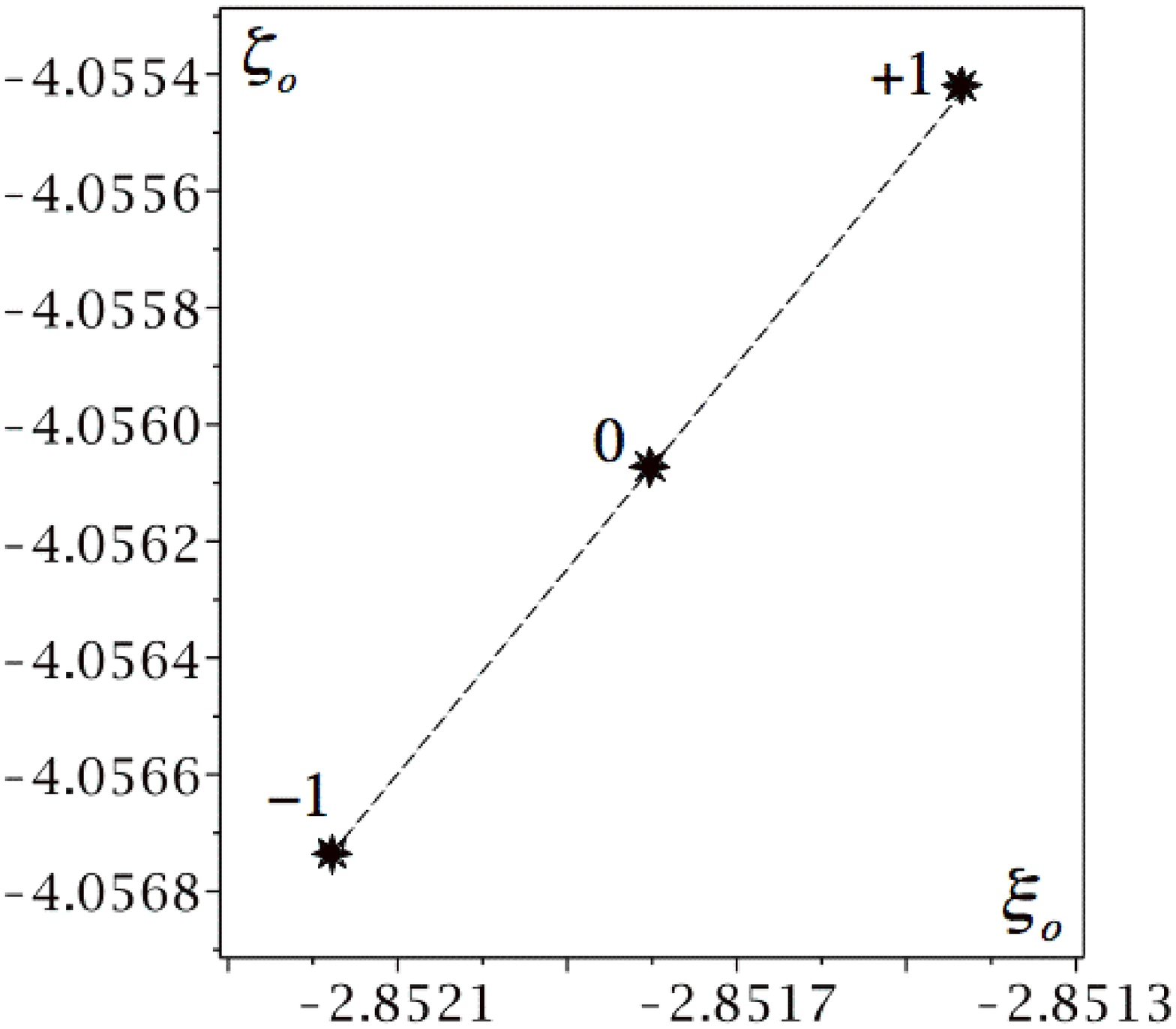}\nn
&&\hspace{3.0cm}({\bf I})\hspace{7.8cm}({\bf II})\nn
&&\hspace{0cm}\includegraphics[width=6cm]{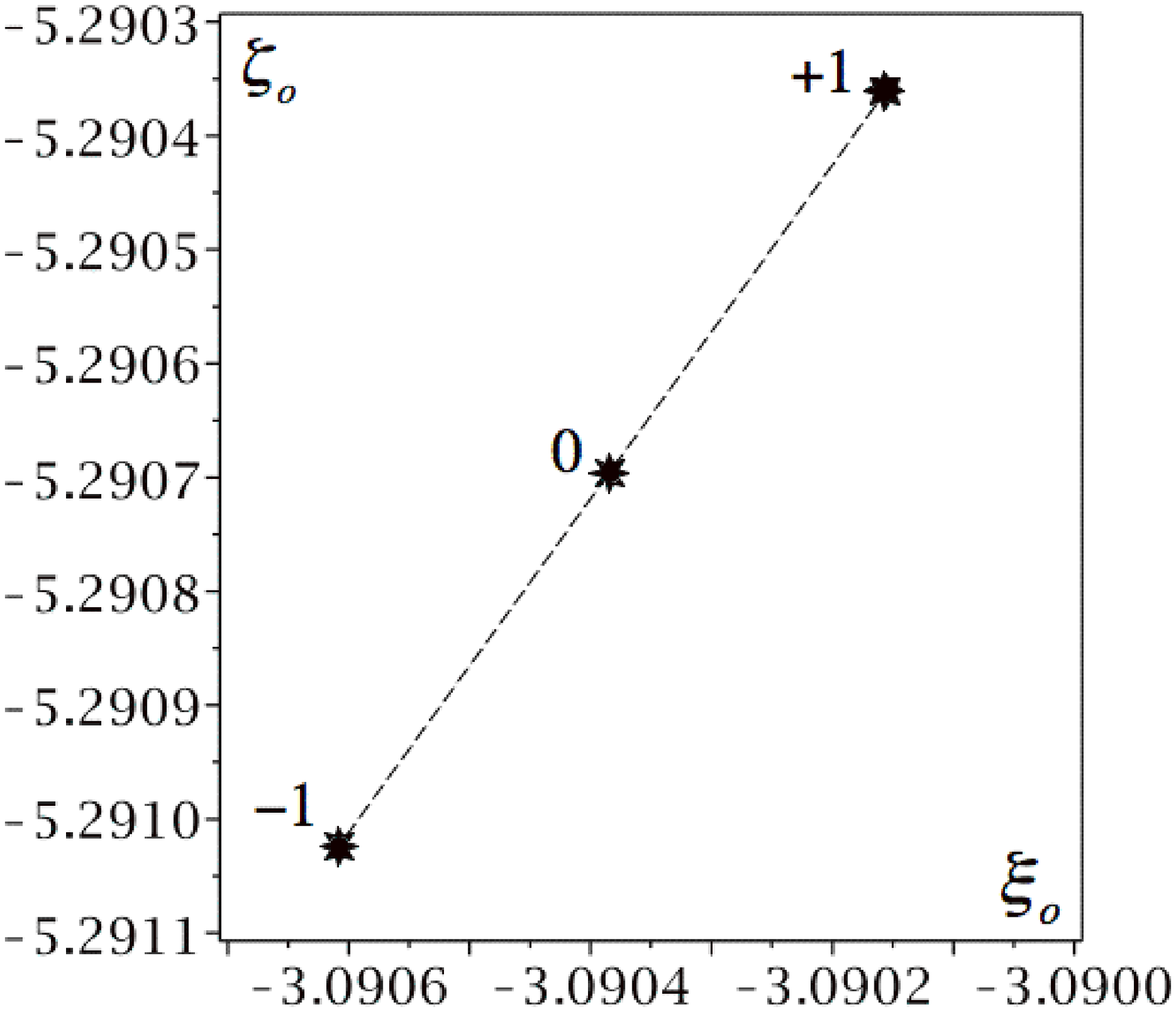}
\hspace{2cm}\includegraphics[width=6cm]{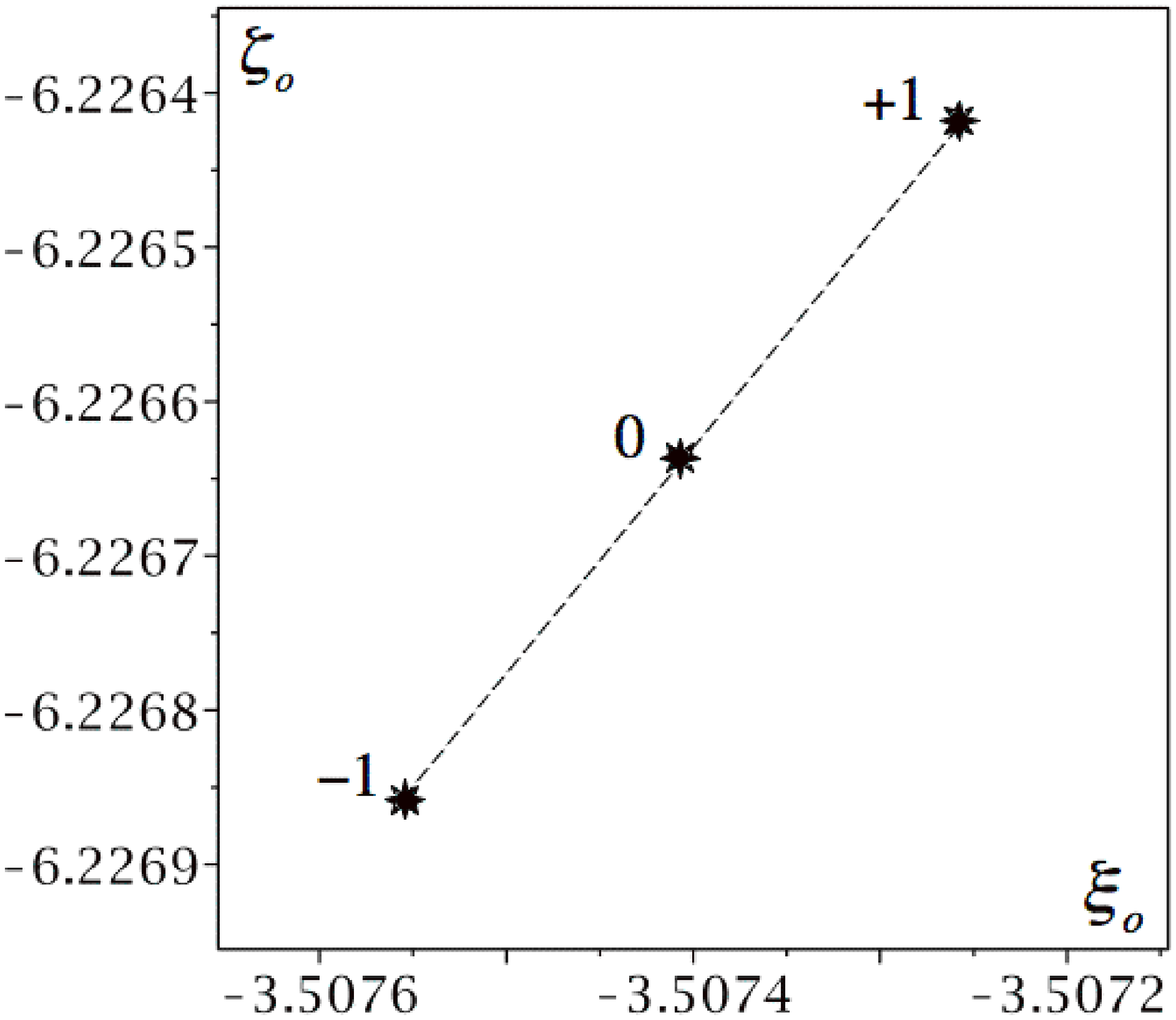}\nn
&&\hspace{3.0cm}({\bf III})\hspace{7.8cm}({\bf IV})\nonumber
\ea
\caption{Images of a point-like emitter. Points $-1$ and $+1$ correspond to the left and right polarized photons, and point $0$ corresponds to a null geodesic.  Figures correspond to the same initial direction of the emitter,  $(y_{e}=\cos\theta_{e}, \phi_{e})=(\sqrt{2}/2,0)$ and different positions of the observer $(y_{o}=\cos\theta_{o}, \phi_{o})$: $(0.85,7.0)$, $\sigma_{yo}=-1$ for Fig. ({\bf I}); $(0.4,4.5)$, $\sigma_{yo}=1$ for Fig. ({\bf II}); $(0.05,4.0)$, $\sigma_{yo}=1$ for Fig. ({\bf III}); $(-0.12,3.8)$, $\sigma_{yo}=1$ for Fig. ({\bf IV}).}\label{F8}
\end{center}
\end{figure*}
\begin{table*}[htdp]
\caption{Numerical data for Fig. \ref{F8}.}
\begin{center}
\begin{tabular}{|c|c|c|c|c|c|c|c|c|c|}
\hline
Point& $\ve$ & $p_{e}$, plot ({\bf I})& $q_{e}$, plot ({\bf I})& $p_{e}$, plot ({\bf II})& $q_{e}$, plot ({\bf II})& $p_{e}$, plot ({\bf III})& $q_{e}$, plot ({\bf III})& $p_{e}$, plot ({\bf IV})& $q_{e}$, plot ({\bf IV})\\ \hline
$-1$& $-0.01$& $1.978342$& $11.739188$& $2.614062$& $17.598726$& $3.086741$& $28.016331$& $3.482207$& $38.936516$\\
0& $0$& $1.980668$& $11.721593$& $2.613719$& $17.593052$& $3.086517$& $28.012854$& $3.482061$& $38.933788$\\
+1& $+0.01$& $1.982997$& $11.703334$& $2.613382$& $17.587374$& $3.086292$& $28.009333$& $3.481913$& $38.931049$\\ \hline
\end{tabular}
\end{center}
\label{T1}
\end{table*}

The situation is different for motion of circularly polarized photons. Because of the axial symmetry, $p$ remains an integral of motion, while$q$ is not, $q_o\ne q_e$. To analyze how the parameter $q$ changes, we introduce the following function $q(\ell)$:
\ba\n{ql}
q(\ell)&=&\frac{(dy/d\phi)^{2}}{\delta^{2}(1-y^{2})^{3}}[p\delta+\alpha x(2-\alpha px)(1-y^{2})]^{2}\nn
&+&\frac{p^{2}y^{2}}{(1-y^{2})} -\alpha^{2}y^{2}\,.
\ea
This function is defined by solving the relations \eq{12b} and \eq{12c} for $q$. Certainly, to obtain a similar expression for $q(\ell)$ one can use the other pair of equations, Eqs. \eq{12a} and \eq{12c}. However, the constraint \eq{4} guarantees that these expressions for $q(\ell)$ are identical. The function \eq{ql} coincides with $q_e$ and $q_o$ at ${\cal J}^-$ and ${\cal J}^+$, respectively.

\subsection{Calculation scheme and results}

The calculation of the polarization dependent splitting of images is a two-point boundary value problem.
In our calculations we proceed as follows.
First we chose a null geodesic connecting the points $(\theta_e, \phi_{e}=0)$ and $(\theta_o,\phi_o)$ and determine the corresponding values of the parameters $(p_e, q_e)$, which are constant along the geodesic. To find a null ray connecting the same points for $\ve\ne 0$ we use the shooting method and implement a numerical code written in Fortran (see, e.g., \cite{fortran}).  Namely, we use the initial seed values of $p_e$ and $q_e$ and using the procedure described in Appendix A integrate the dynamical equation \eq{18a} up to the final point. We compare the numerical result for the final angles $(\theta, \phi)$ with the required final data $(\theta_o,\phi_o)$. This comparison allows us to correct the seed values of $p_e$ and $q_e$ and repeat the integration, until the required accuracy is achieved (for more details of the shooting method see, e.g., Chapter 17 of the book \cite{fortran}).
The numerical integration of the dynamical equation uses the fifth-order Runge-Kutta method with the accuracy of $10^{-6}$.

As earlier, we consider an extremal black hole, $\alpha=1$.
To illustrate the results of the calculations, we consider a special case when the inclination angle for incoming photons is $\pi/4$, so that $(y_{e}=\cos\theta_e,\phi_e)=(\sqrt{2}/2,0)$. We also take $\ve=\pm 0.01$. The polarization splitting of the images for different relative positions of the emitter and the observer is illustrated in Figure~\ref{F8}. The  plots show position of the images in the impact plane for the following positions of the observer $(y_{o}=\cos\theta_{o}, \phi_{o})$: $(0.85,7.0)$, $(0.4,4.5)$, $(0.05,4.0)$, and $(-0.12,3.8)$. The calculated values of the parameters $(p_{e},q_{e})$ are given in Table \ref{T1}. The splitting of the images can be characterized by the displacement parameter
\be
\Delta_{o(\pm)}=|\ve|^{-1}\sqrt{ (\Delta \xi_{o(\pm)})^2+ (\Delta \zeta_{o(\pm)})^2}\,,
\ee
where
\be
\Delta \xi_{o(\pm)}\equiv\xi_{o(\pm)}-\xi_{o(0)}\hhh \Delta \zeta_{o(\pm)}\equiv\zeta_{o(\pm)}-\zeta_{o(0)}\,.
\ee
Here $\xi_{o(\pm)}$ and $\zeta_{o(\pm)}$ are the impact parameters of the left $(-)$ and right $(+)$ circularly polarized photons, $\xi_{o(0)}$ and $\zeta_{o(0)}$ correspond to the null geodesic. The values of the displacement  parameter can be found in Table~\ref{T2}.
\begin{table}[htdp]
\caption{Values of the displacement parameter $\Delta_{o(\pm)}$.}
\begin{center}
\begin{tabular}{|c|c|c|}
\hline
\,\,\,\,\,\,Plot\,\,\,\,\,\,& $\,\,\,\,\,\,\Delta_{o(-)}\,\,\,\,\,\,$ & $\,\,\,\,\,\,\Delta_{o(+)}\,\,\,\,\,\,$\\ \hline
({\bf I})& $1.4607$& $1.4831$\\
({\bf II})& $0.0756$& $0.0754$\\
({\bf III})& $0.0398$& $0.0402$\\
({\bf IV})& $0.0263$& $0.0265$\\
\hline
\end{tabular}
\end{center}
\label{T2}
\end{table}

\begin{figure}[htb]
\begin{center}
\includegraphics[width=5cm]{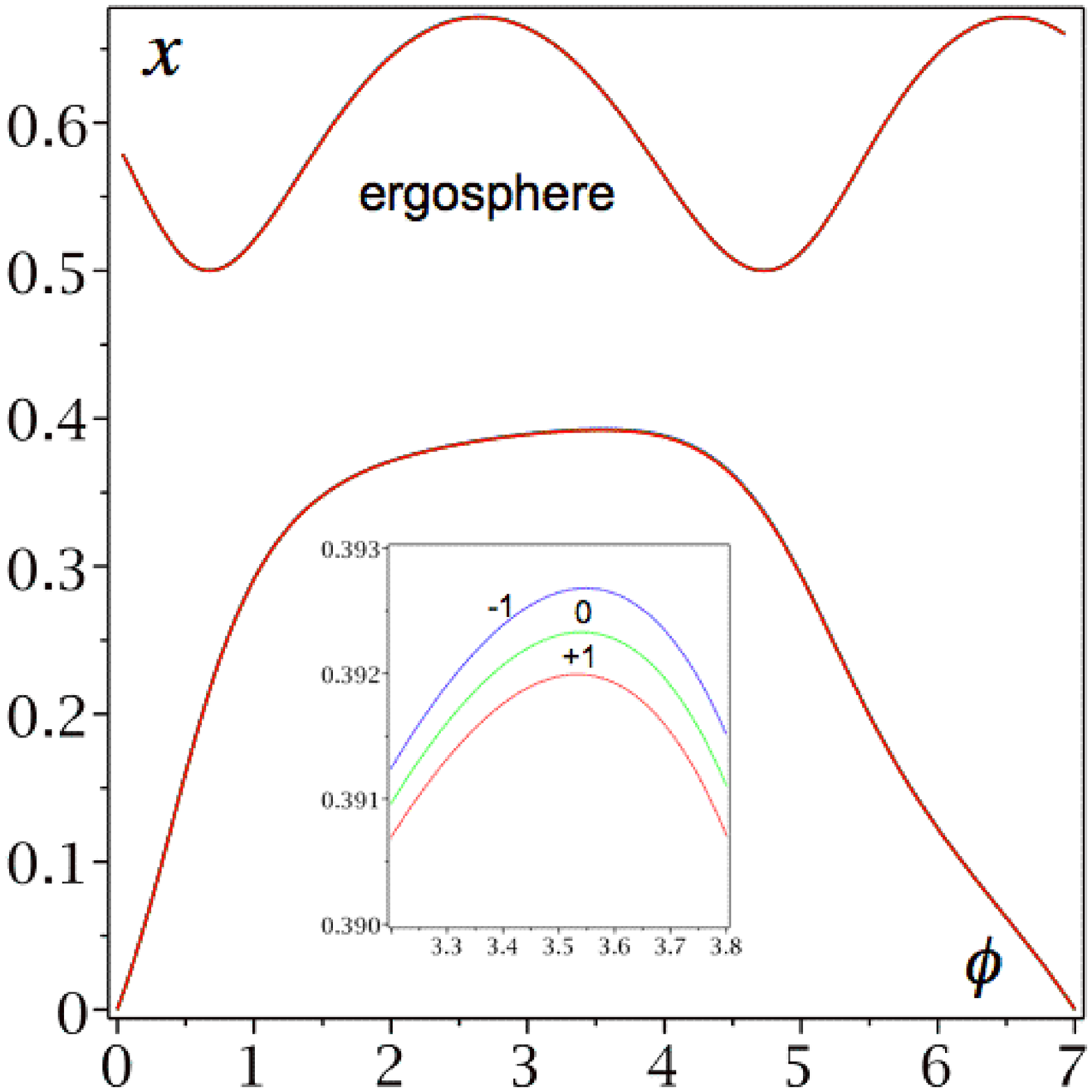}
\caption{Plot $x=x(\phi)$. Curves $-1$ and $+1$ illustrate trajectories of the left and right circularly polarized photons, respectively. Curve $0$ illustrates a null geodesic.}\label{F5}
\end{center}
\end{figure}

\begin{figure}[htb]
\begin{center}
\includegraphics[width=5cm]{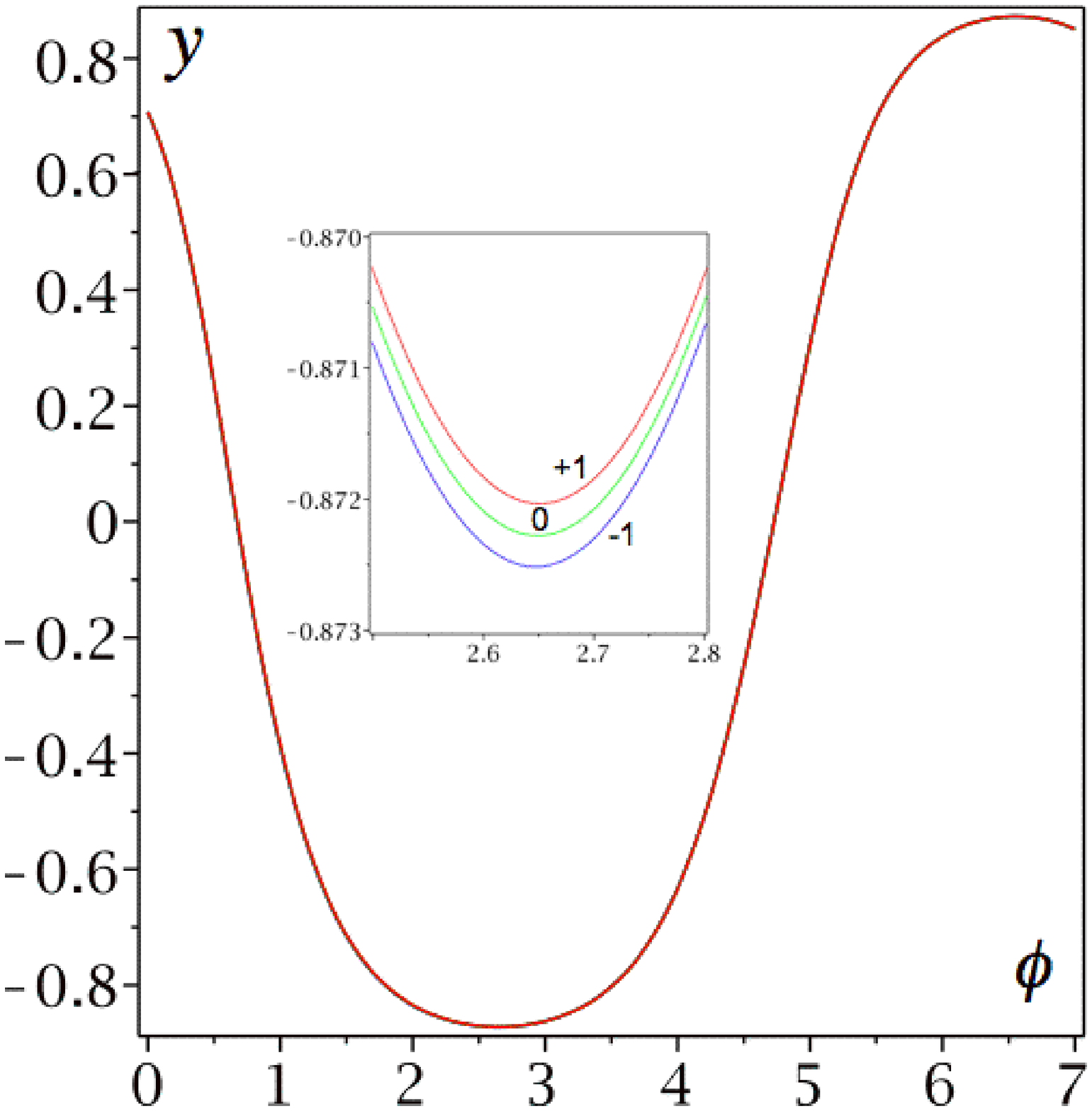}
\caption{Plot $y=y(\phi)$. Curves $-1$ and $+1$ illustrate trajectories of the left and right circularly polarized photons, respectively. Curve $0$ illustrates a null geodesic.}\label{F6}
\end{center}
\end{figure}

\begin{figure}[htb]
\begin{center}
\includegraphics[width=5cm]{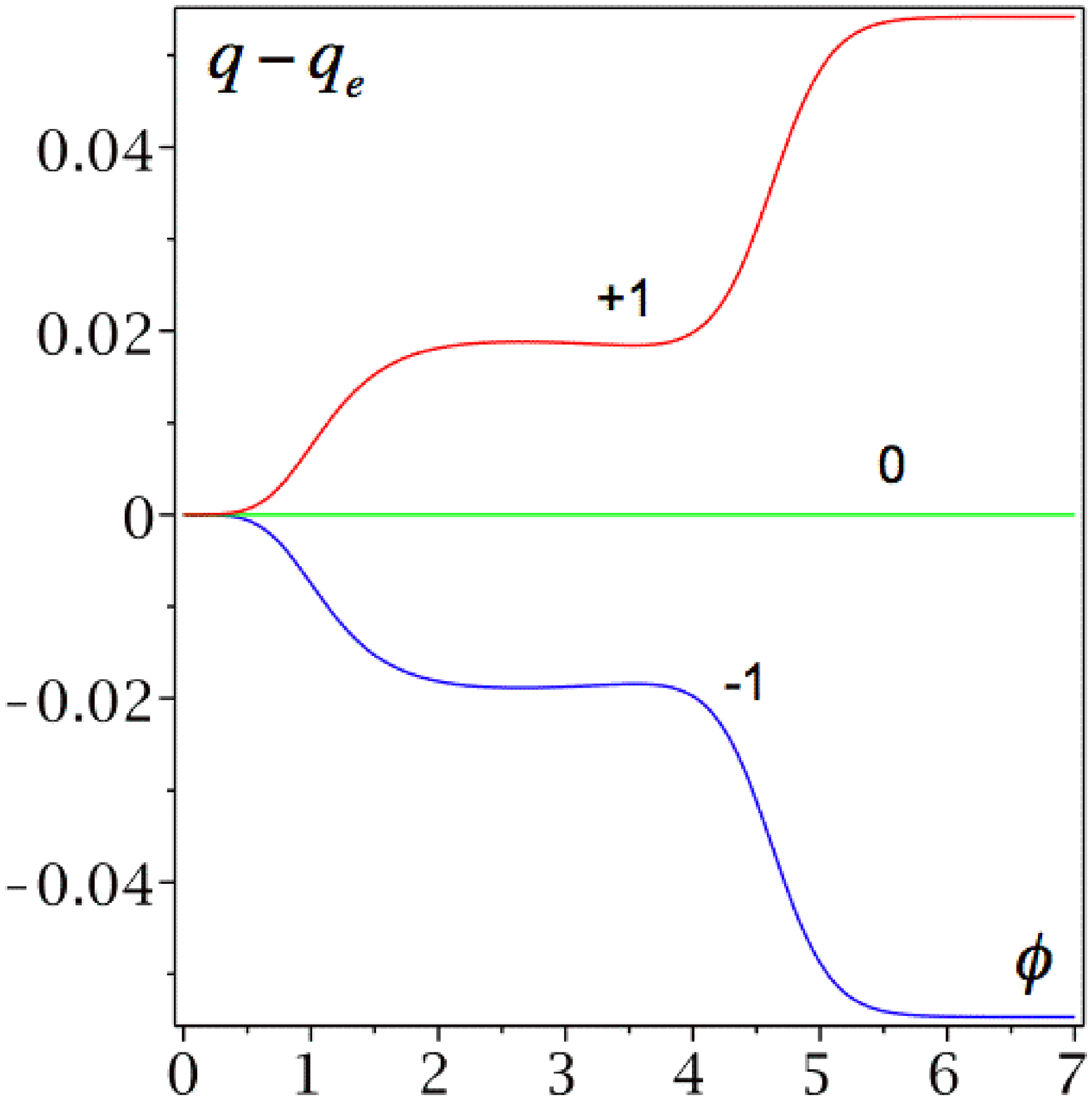}
\caption{Plot $q-q_{e}$ versus $\phi$. Curves $-1$ and $+1$ correspond to the left and right circularly polarized photons, respectively. Curve $0$ corresponds to a null geodesic.}\label{F7}
\end{center}
\end{figure}
Let us discuss photon trajectories for the same position of the emitter $(\sqrt{2}/2,0)$ and for the case when the observer is located at the point $(0.85,7.0)$. We chose this case only for an illustration. Let us emphasize that in this case, the bending angle is larger than $\pi$, that is, the photon passes the black hole at a relatively short distance. This makes the polarization splitting of images more pronounced. However, the linear in $\ve$ approximation remains valid. Plots presented in Figures~\ref{F5}--\ref{F6} show the photon coordinates $x$ and $y$ as functions of the angle $\phi$. In Figure~\ref{F5} the lower solid line demonstrates how photons starting at ${\cal J}^-$ ($x=0$) come close to the black hole, reach their turning point, where $x$ takes its maximal value, and propagate to ${\cal J}^+$ ($x=0$). The upper solid curve shows position of the ergosphere with respect to a photon. To define the position of the ergosphere one uses Eq. \eq{horerg}, where $y=y(\phi)$ is defined by the photon trajectory. We  calculated photon trajectories for a geodesic case (curve 0), and for the left ($\ve=-0.01$) and the right ($\ve=+0.01$) polarized photons, (curves $-1$ and $+1$). The difference between these curves is small. To demonstrate this, we presented a zoomed part of the trajectories near their radial turning point. Figure~\ref{F6} contains similar information about $y=y(\phi)$.  The numerical data for the asymptotic parameters $(p_e,q_e)$ are given in Table~\ref{T3}. 
\begin{table}[htdp]
\caption{Numerical data for Figs.  \ref{F5}--\ref{F7}.}
\begin{center}
\begin{tabular}{|c|c|c|c|}
\hline
Curve& $\ve$ & $p_{e}$ & $q_{e}$\\ \hline
$-1$& $-0.01$& $1.978342$& $11.739188$\\
0& $0$& $1.980668$& $11.721593$\\
+1& $+0.01$& $1.982997$& $11.703334$\\ \hline
\end{tabular}
\end{center}
\label{T3}
\end{table}

As we explained earlier, $p$ is an exact integral of motion, while $q$ changes along trajectory of a circularly polarized photon [see Eq. \eq{ql}]. Figure~\ref{F7} illustrates this change.
To estimate an error in the linear in $\ve$ approximation we introduce the following quantities:
\be\n{pqe}
\Delta p_{(\pm)}=p(\pm|\ve|)-p(0)\hhh\Delta q_{(\pm)}=q(\pm|\ve|)-q(0)\,,
\ee
and define the relative errors of the first order expansion in $\ve$
\be
\delta_{p_{(\pm)}}=\frac{|\Delta p_{(+)}+\Delta p_{(-)}|}{2|\Delta p_{(\pm)}|}\hhh\delta_{q_{(\pm)}}=\frac{|\Delta q_{(+)}+\Delta q_{(-)}|}{2|\Delta q_{(\pm)}|}\,.
\ee
Using the data presented in Table \ref{T3}, we derive $\delta_{p_{(-)}}\approx0.07\%$,  $\delta_{p_{(+)}}\approx0.06\%$, $\delta_{q_{(-)}}\approx1.89\%$, and  $\delta_{q_{(+)}}\approx1.82\%$. This means that non-linear effects are small and can be neglected.

\subsection{Arrival time}

Let us briefly discuss the last of the three problems formulated at the end of Sec.~IV. Namely, how the arrival time depends on the photon polarization. In order to determine this quantity, one can use the relation \eq{time}, where the integration is performed along a photon trajectory. If one parameterizes the photon trajectory by the star time parameters, as it is done in our calculations (see Appendix A), the quantity
\be\n{w0}
\Delta w=w_{+}|_{{\cal J}^{+}}-w_{-}|_{{\cal J}^{-}}\,
\ee
is obtained simply as the difference between the initial and the final values of the corresponding star time parameters. Let us again emphasize that the value of $\Delta w$ depends on the choice of the upper limit in the integral in \eq{time}, which we chose to be equal to $1/2$, that corresponds to $r=2M$. However, when one compares the registration times of photons of different polarizations emitted simultaneously, the difference in their arrival times is uniquely determined. One has\be
\Delta \tau_{(\pm)}=\Delta w_{(\pm)}-\Delta w_{(0)}\, .
\ee

Here we present the results of the time delay of the polarized photons for the one of the above cases, when the spin-optical effects are more pronounced. The related data are given in Table \ref{T3}. As before, we use the subscript $(0)$ for a null geodesic, while the subscripts $(-)$ and $(+)$ stand for the left and right polarization of photons, respectively. For the null geodesic, we have the following value of the parameter \eq{w0}:
\be
\Delta w_{(0)}=13.458554\,.
\ee
For the above data one has
\be
\Delta \tau_{(-)}/\ve^{2}\approx0.17\hhh \Delta \tau_{(+)}/\ve^{2}\approx0.18\,.
\ee
These relations imply that polarized photons arrive later than `non-polarized' ones, and that the time delay is the effect of the second order in the parameter $\ve$. Both the conclusions are in agreement with the general results on the time delay discussed at the end of Sec. II.

\section{Discussion}

In this paper we applied the modified geometric optics approach developed in \cite{FS} to the problem of scattering of polarized light by a rotating black hole. We demonstrated that trajectories of polarized photons are null curves, which coincide with null geodesics only in the limit $\omega\to\infty$. A deviation of the null rays from null geodesics is controlled by the small parameter $\ve=\pm(2\omega M)^{-1}$. Since the polarization dependent term in the photon propagation equations \eq{18a}--\eq{18c} rapidly falls down at infinity, the initial ${\cal D}_{e}$ and the  final ${\cal D}_{o}$ scattering data for null rays are the same as those for null geodesics. These data include the (null) time and direction at ${\cal J}^{\pm}$, as well as a vector in the 2D space of impact parameters. Working in (3+1) formalism we studied the equations for trajectories of polarized photons in the Kerr geometry. We developed a solver, which allows one to integrate the required equations and to solve the scattering problem, that is, to find numerically the scattering operator $\Psi(\ve): {\cal D}_{e}\to{\cal D}_{o}$.
We used this approach to study how bending angles of the scattered photons depend on their polarization. We also analyzed how the splitting of the image of a point-like source depends on polarization and calculated the time delay for polarized photons.
It should be emphasized that the photon polarization enters the dynamical equations through the factor $\ve$, which depends on the photon frequency. This means that all the above discussed effects are frequency dependent.

Unfortunately, in the framework of the approach used in the paper, one cannot consider effects for polarized photons passing quite close to a rotating black hole, e.g., propagating through its ergosphere. This happens because the calculations are based on the (3+1) split of the spacetime obtained by projecting it along Killing trajectories. This split does not work on the ergosurface and inside of it. In principle, one can avoid this problem, for example, by using another Killing vector $\tilde{\BM{\xi}}=\BM{\xi}_{(t)}+\Omega \BM{\xi}_{(\phi)}$, with properly chosen constant $\Omega$, so that the new Killing vector is well defined inside the ergosphere. After that, one can glue the trajectories corresponding to the modified GO approach obtained with the help of $\tilde{\BM{\xi}}$ and $\BM{\xi}_{(t)}$ in the spacetime domain where both of the vectors are timelike. Another, more interesting problem is to develop the  modified GO approach directly in a four dimensional spacetime. Such approach would allow one to study the polarization dependent effects for photons which closely approach a rotating black hole. In particular, it is interesting to study how polarization of light modifies the black hole shadow. One can expect that 
the black hole shadow will in fact be smeared for the polarized light, and if the `bright radiation' behind the rotating black hole is not monochromatic, the position of the shadow rim will depend on the frequency of the radiation. In other words, one would observe a rainbow effect for the black hole shadow.

The polarization splitting of images might, in principle, be detectable by astronomical observations. However, this depends on sensitivity of the measuring devices. It is interesting to study possible applications of the obtained results to real astrophysical problems. 
 
\appendix

\section{Details of the numerical computation}

Here we present the main steps of the numerical computation. Equations (\ref{18a})--(\ref{18c}) determine the coordinates of a null curve $(x,y,\phi, \tau)$ as functions of the proper length $\ell$. Direct integration of the equations in this form is not convenient, since for the scattering problem $\ell$ changes in an infinite interval. To deal with finite quantities parameterizing the entire null curve (from ${\cal J}^-$ to ${\cal J}^+$), we use the star time coordinates $w_{\pm}$. Namely, we use $w_-$ for the part of the null curve from ${\cal J}^-$ to its radial turning point, and $w_+$ from this point to ${\cal J}^+$. The condition \eq{wpm} uniquely relates these parameters at the turning point.

The complete set of the equations for $x$, $y$, and $\phi$ as functions of $w_{\pm}$ written in the first order form is
\ba\n{sys}
\frac{dx}{dw_{\pm}}&=&\frac{X}{A}\hhh\frac{dy}{dw_{\pm}}=\frac{Y}{A}\,,\nn
\frac{dX}{dw_{\pm}}&=&-\frac{1}{2A\gamma_{xx}}(\gamma_{xx,x}X^{2}+2\gamma_{xx,y}XY-\gamma_{yy,x}Y^{2})\nn
&+&\frac{\gamma_{\phi\phi,x}}{2A\gamma_{xx}\gamma_{\phi\phi}^{2}}(p-\gamma_{\phi})^{2}-\frac{(p-\gamma_{\phi})}{A\sqrt{\gamma}}(\mbox{curl}\,\BM{\gamma})_{y}\nn
&+&\ve\frac{\gamma_{yy}(\mbox{curl}\,\mbox{curl}\BM{\gamma})_{\phi}}{A\sqrt{\gamma}}Y\,,\nn
\frac{dY}{dw_{\pm}}&=&-\frac{1}{2A\gamma_{yy}}(\gamma_{yy,y}Y^{2}+2\gamma_{yy,x}XY-\gamma_{xx,y}X^{2})\nn
&+&\frac{\gamma_{\phi\phi,y}}{2A\gamma_{yy}\gamma_{\phi\phi}^{2}}(p-\gamma_{\phi})^{2}+\frac{(p-\gamma_{\phi})}{A\sqrt{\gamma}}(\mbox{curl}\,\BM{\gamma})_{x}\nn
&-&\ve\frac{\gamma_{xx}(\mbox{curl}\,\mbox{curl}\BM{\gamma})_{\phi}}{A\sqrt{\gamma}}X\,,\nn
\frac{d\phi}{dw_{\pm}}&=&\frac{(p-\gamma_{\phi})}{A\gamma_{\phi\phi}}\,,
\ea
where
\be
A=1+\frac{\gamma_{\phi}(p-\gamma_{\phi})}{\gamma_{\phi\phi}}-\frac{\eta}{x^{2}\delta}|X|\,.
\ee
The constraint \eq{4}, which takes the form
\be\n{ConA}
\gamma_{xx}X^{2}+\gamma_{yy}Y^{2}+\frac{(p-\gamma_{\phi})^{2}}{\gamma_{\phi\phi}}=1\, ,
\ee
is used to control the accuracy of the calculations.

In principle, one can use these equations to start the integration from a point with $x=0$ corresponding to ${\cal J}^-$. In fact, we found that it is more convenient to use a slightly modified approach, which allows one to get higher precision. Namely, in the vicinity of ${\cal J}^{\pm}$ we parameterize the coordinates $x$ and $y$ by the angle $\phi$, where for a rotating black hole this parameter is monotonic along null curves. For this parametrization the initial values for a null curve at ${\cal J}^-$ are
\ba
&&x|_{\phi=0}=0\hh y|_{\phi=0}=y_e\,,\nn
&&\left. \frac{dx}{d\phi}\right|_{\phi=0}=\frac{(1-y_{e}^{2})}{p}\, ,\n{27}\\
&&\left. \frac{dy}{d\phi}\right|_{\phi=0}=\sigma_{y}\frac{(1-y_{e}^{2})}{p}\sqrt{(q+\alpha^{2}y_{e}^{2})(1-y_{e}^{2})
-p^{2}y_{e}^{2}}\,.\nonumber
\ea
The asymptotic integrals of motion $p$ and $q$, which enter these relations are related to the impact parameters by means of the expressions \eq{Lz} and \eq{qqq}.

The equation for $x(\phi)$ is of the form
\be\n{24}
\frac{d^{2}x}{d\phi^{2}}={\cal A}\left(\frac{dx}{d\phi}\right)^{2}+{\cal B}\frac{dx}{d\phi}\frac{dy}{d\phi}
+{\cal C}\left(\frac{dy}{d\phi}\right)^{2}+{\cal D}\frac{dy}{d\phi}+{\cal E}\,,
\ee
where
\ba
{\cal A}&=&\frac{\gamma_{\phi\phi,x}}{\gamma_{\phi\phi}}-\frac{\gamma_{xx,x}}{2\gamma_{xx}}-
\frac{\gamma_{xx}\gamma_{\phi\phi}}{(p-\gamma_{\phi})\sqrt{\gamma}}(\mbox{curl}\,\BM{\gamma})_{y}\,,\nn
{\cal B}&=&\frac{\gamma_{\phi\phi,y}}{\gamma_{\phi\phi}}-\frac{\gamma_{xx,y}}{\gamma_{xx}}+
\frac{\gamma_{yy}\gamma_{\phi\phi}}{(p-\gamma_{\phi})\sqrt{\gamma}}(\mbox{curl}\,\BM{\gamma})_{x}\,,\n{25}\\
{\cal C}&=&\frac{\gamma_{yy,x}}{2\gamma_{xx}}\hhh
{\cal D}=\ve\frac{\gamma_{yy}\gamma_{\phi\phi}}{(p-\gamma_{\phi})\sqrt{\gamma}}(\mbox{curl}\,
\mbox{curl}\,\BM{\gamma})_{\phi}\,,\nn
{\cal E}&=&\frac{\gamma_{\phi\phi,x}}{2\gamma_{xx}}-\frac{\gamma^{2}_{\phi\phi}}{(p-\gamma_{\phi})
\sqrt{\gamma}}(\mbox{curl}\,\BM{\gamma})_{y}\,.\nonumber
\ea
A similar equation can be written for $y(\phi)$. This equation is obtained if one makes an interchange $x\leftrightarrow y$ in all the expressions that enter \eq{24} and \eq{25} and reverse the sign of the terms containing $\mbox{curl}\,\BM{\gamma}$ and $\mbox{curl}\,\mbox{curl}\,\BM{\gamma}$. To solve these equations we
use the seventh-eighth order Runge-Kutta method with the accuracy of $10^{-9}$ and check our computation by using the constraint \eq{ConA} adapted to this form of equations. We integrate the equations from the point where $x=0$ up to the point where $x=x_{1}$ is of the order of $10^{-2}$. Since one cannot guarantee that $\phi$ is a monotonic coordinate along the entire null curve, we use the result obtained at the point $x_{1}$ to find the initial conditions for the system \eq{sys}. Then we integrate this system by using the seventh-eighth order Runge-Kutta method with the accuracy of $10^{-9}$ and check our computation by using the constraint \eq{ConA}. We integrate the equations through a radial turning point, until we reach the point where $x=x_{2}$ is of the order of $10^{-2}$.
At this point we derive new initial conditions and use them to continue the calculations by integrating the system \eq{24} up to the end point where $x\approx10^{-8}$. This allows us to obtain with the required accuracy the  angular coordinates $(\theta_{o},\phi_{o})$ at ${\cal J}^{+}$ for the outgoing null rays.

\begin{acknowledgments}

This work was partly supported  by  the Natural Sciences and Engineering Research Council of Canada. The authors are grateful to the Killam Trust for its financial support.

\end{acknowledgments}

\end{document}